\documentclass [12pt]{article}
\usepackage {graphicx}
\usepackage{bm}
\usepackage{booktabs}
\usepackage{multirow}
\usepackage{amsmath}
\usepackage{amssymb}
\usepackage{cite}

\renewcommand{\baselinestretch}{1.3}

\newtheorem{model ass}[theo]{Model Assumptions}

\begin{document}

\title{Implementing Loss Distribution Approach for Operational Risk}

\author{Pavel V.~Shevchenko \\ \footnotesize{CSIRO Mathematical and Information
Sciences, Sydney,} \\ \footnotesize{Locked Bag 17, North Ryde, NSW,
1670, Australia.}\\ \footnotesize{e-mail:
Pavel.Shevchenko@csiro.au}}

\date{1st version: 21 October 2008\\
This version: 26 July 2009}

\maketitle

\begin{abstract}
\noindent To quantify the operational risk capital charge under the
current regulatory framework for banking supervision, referred to as
Basel II, many banks adopt the Loss Distribution Approach. There are
many modeling issues that should be resolved to use the approach in
practice. In this paper we review the quantitative methods suggested
in literature for implementation of the approach. In particular, the
use of the Bayesian inference method that allows to take expert
judgement and parameter uncertainty into account, modeling
dependence and inclusion of insurance are discussed.

\vspace{1cm} \noindent \textbf{Keywords:} operational risk; loss
distribution approach; Bayesian inference; dependence modeling;
Basel II.
\end{abstract}

\pagebreak

\section{Operational risk under Basel II}
\label{intro_sec} Under the current regulatory framework for the
banking industry \cite{BaselII06}, referred to as Basel II, the
banks are required to hold adequate capital against operational risk
(OR) losses. OR is a new category of risk, in addition to market and
credit risks, attracting capital charge and defined by Basel II
\cite[p.144]{BaselII06} as: ``\emph{[\ldots] the risk of loss
resulting from inadequate or failed internal processes, people and
systems or from external events. This definition includes legal
risk, but excludes strategic and reputational risk.}'' Similar
regulatory requirements for the insurance industry are referred to
as Solvency 2. OR is significant in many financial institutions.
Examples of extremely large OR losses are: Barings Bank (loss GBP
1.3 billion in 1995), Sumitomo Corporation (loss USD 2.6 billion in
1996), Enron (USD 2.2 billion in 2001), and recent loss in
Soci\'{e}t\'{e} G\'{e}n\'{e}rale (Euro 4.9 billion in 2008). In
Basel II, three approaches can be used to quantify the OR annual
capital charge $C$, see \cite[pp.144-148]{BaselII06}:

\begin{itemize}
\item{The Basic Indicator Approach: $C = \alpha \frac{1}{n} \sum\nolimits_{j = 1}^n {GI(j)} $, $\alpha = 0.15$, where
$GI(j)$, $j = 1,..,n$ are the annual positive gross incomes over the
previous three years.}

\item{The Standardised Approach: $C = \frac{1}{3}\sum\nolimits_{j = 1}^3 {\max [\sum\nolimits_{i = 1}^8 {\beta _i
GI_i (j)} ,0]} $, where $\beta _i $, $i = 1,\ldots,8$ are the
factors for eight business lines (BL) listed in Table
\ref{BURT_table} and $GI_i (j)$, $j = 1,2,3$ are the annual gross
incomes of the $i$-th BL in the previous three years.}

\item{The Advanced Measurement Approaches (AMA): a bank can calculate
the capital charge using internally developed model subject to
regulatory approval.}

\end{itemize}

\noindent A bank intending to use the AMA should demonstrate
accuracy of the internal models within the Basel II risk cells
(eight business lines times seven risk types, see Table
\ref{BURT_table}) relevant to the bank and satisfy some criteria,
see \cite[pp.148-156]{BaselII06}, including:

\begin{itemize}
\item{The use of the internal data, relevant external data, scenario
analysis and factors reflecting the business environment and
internal control systems;}

\item{The risk measure used for capital charge should correspond to
the 99.9{\%} confidence level for a one-year holding period;}

\item{Diversification benefits are allowed if dependence modeling is
approved by a regulator;}

\item{Capital reduction due to insurance is capped by 20{\%}.}
\end{itemize}

\noindent A popular method under the AMA is the loss distribution
approach (LDA). Under the LDA, banks quantify distributions for
frequency and severity of OR losses for each risk cell (business
line/event type) over a one-year time horizon. The banks can use
their own risk cell structure but must be able to map the losses to
the Basel II risk cells. There are various quantitative aspects of
the LDA modeling discussed in several books
\cite{King01,Cruz02,Cruz04,Panjer06, McFrEm05, ChRaFa07} and various
papers, e.g. \cite{ChEmNe06,FrMoRo04,AuKl06} to mention a few. The
commonly used LDA model for calculating the total annual loss
$Z_{(\bullet )} (t)$ in a bank (occurring in the years $t =
1,2,\ldots)$ can be formulated as

\begin{equation}
\label{LDAmodel_eq} Z_{(\bullet )} (t) = \sum\limits_{j = 1}^J {Z_j
(t)} ;\quad Z_j (t) = \sum\limits_{i = 1}^{N_j (t)} {X_j^{(i)} (t)}
.
\end{equation}

\noindent Here, the annual loss $Z_j (t)$ in risk cell $j$ is
modeled as a compound process over one year with the frequency
(annual number of events) $N_j (t)$ implied by a counting process
(e.g. Poisson process) and random severities $X_j^{(i)} (t)$, $i =
1,\ldots,N_j (t)$. Typically, the frequencies and severities are
assumed independent. Estimation of the annual loss distribution by
modeling frequency and severity of losses is a well-known actuarial
technique, see e.g. Klugman \textit{et al}.\cite{KlPaWi98}. It is
also used to model solvency requirements for the insurance industry,
see e.g. Sandstr\"{o}m \cite{Sandstrom06} and W\"{u}thrich and Merz
\cite{WuMe08}. Under the model (\ref{LDAmodel_eq}), the capital is
defined as the 0.999 Value at Risk (VaR) which is the quantile of
the distribution for the next year annual loss $Z_{(\bullet )} (T +
1)$:

\begin{equation}
\label{VaRdef_eq} VaR_q (Z_{(\bullet )} (T + 1)) = F_{Z_{(\bullet )}
(T + 1)}^{ - 1} (q) = \inf \{z:\Pr [Z_{(\bullet )} (T + 1) > z] \le
1 - q\}
\end{equation}

\noindent at the level $q = 0.999$. Here, index $T$+1 refers to the
next year and notation $F_Y^{ - 1} (q)$ denotes the inverse
distribution of a random variable (rv) $Y$. The capital can be
calculated as the difference between the 0.999 VaR and expected loss
if the bank can demonstrate that the expected loss is adequately
captured through other provisions. If correlation assumptions can
not be validated between some groups of risks (e.g. between business
lines) then the capital should be calculated as the sum of the 0.999
VaRs over these groups. This is equivalent to the assumption of
perfect positive dependence between annual losses of these groups.

In this paper, we review some methods proposed in the literature for
the LDA model (\ref{LDAmodel_eq}). In particular, we consider the
Bayesian inference approach that allows to account for expert
judgment and parameter uncertainty which are important issues in
operational risk management.

The paper is organised as follows. Section \ref{data_sec} describes
the requirements for the data that should be collected and used for
Basel II AMA. Section \ref{truncData_sec} and Section
\ref{sevTail_sec} are focused on modeling truncated data and the
tail of severity distribution respectively. Calculation of compound
distributions is considered in Section \ref{CompDFcalc_sec}. In
Section \ref{ModelEstimation_sec}, we review the estimation of the
frequency and severity distributions using frequentist and Bayesian
inference approaches. Combining different data sources (internal
data, expert judgement and external data) is considered in Section
\ref{DataCombining_sec}. Modeling dependence and insurance are
discussed in Section \ref{Dependence_sec} and Section
\ref{Insurance_sec} respectively. Finally, Section
\ref{CapitalViaPredDF_sec} presents the estimation of the capital
charge via full predictive distribution accounting for parameter
uncertainty.

\section{Data}
\label{data_sec} Basel II specifies requirement for the data that
should be collected and used for AMA. In brief, a bank should have:
internal data, external data and expert opinion data. In addition,
internal control indicators and factors affecting the businesses
should be used. Development and maintenance of OR databases is a
difficult and challenging task. Some of the main features of the
required data are summarized as follows.

~

\noindent\textbf{Internal data.} The internal data should be
collected over a minimum five year period to be used for capital
charge calculations (when the bank starts the AMA, a three-year
period is acceptable). Due to a short observation period, typically,
the internal data for many risk cells contain few (or none) high
impact low frequency losses. A bank must be able to map its
historical internal loss data into the relevant Basel II risk cells
in Table \ref{BURT_table}. The data must capture all material
activities and exposures from all appropriate sub-systems and
geographic locations. A bank can have an appropriate reporting
threshold for internal data collection, typically of the order of
Euro 10,000. Aside from information on gross loss amounts, a bank
should collect information about the date of the event, any
recoveries of gross loss amounts, as well as some descriptive
information about the drivers of the loss event.

~

\noindent\textbf{External data.} A bank's OR measurement system must
use relevant external data. These data should include data on actual
loss amounts, information on the scale of business operations where
the event occurred, and information on the causes and circumstances
of the loss events. Industry data are available through external
databases from vendors (e.g. Algo OpData provides publicly reported
OR losses above USD 1 million) and consortia of banks (e.g. ORX
provides OR losses above Euro 20,000 reported by ORX members). The
external data are difficult to use directly due to different volumes
and other factors. Moreover, the data have a survival bias as
typically the data of all collapsed companies are not available.
Several Loss Data Collection Exercises (LDCE) for historical OR
losses over many institutions were conducted and their analyses
reported in the literature. In this respect, two papers are of high
importance: Moscadelli \cite{Moscadelli04} analysing 2002 LDCE and
Dutta and Perry \cite{DuPe06} analysing 2004 LDCE where the data
were mainly above Euro 10,000 and USD 10,000 respectively.

~

\noindent\textbf{Scenario Analysis/expert opinion.} A bank must use
scenario analysis in conjunction with external data to evaluate its
exposure to high-severity events. Scenario analysis is a process
undertaken by experienced business managers and risk management
experts to identify risks, analyse past internal/external events,
consider current and planned controls in the banks; etc. It may
involve: workshops to identify weaknesses, strengths and other
factors; opinions on the impact and likelihood of losses; opinions
on sample characteristics or distribution parameters of the
potential losses. As a result some rough quantitative assessment of
risk frequency and severity distributions can be obtained. Scenario
analysis is very subjective and should be combined with the actual
loss data. In addition, it should be used for stress testing, e.g.
to assess the impact of potential losses arising from multiple
simultaneous loss events.

~

\noindent\textbf{Business environment and internal control factors.}
A bank's methodology must capture key business environment and
internal control factors affecting OR. These factors should help to
make forward-looking estimation, account for the quality of the
controls and operating environments, and align capital assessments
with risk management objectives.

\section{A note on modeling truncated data}
\label{truncData_sec} As mentioned above, typically internal data
are collected above some low level of the order of Euro 10,000.
Generally speaking, omitting data increases uncertainty in modeling
but having a reporting threshold helps to avoid difficulties with
collection of too many small losses. Often, the data below a
reported level are simply ignored in the analysis, arguing that the
capital is mainly determined by the low frequency heavy tailed
severity risks. However, the impact of data truncation for other
risks can be significant. Even if the impact is small often it
should be estimated to justify the reporting level. Recent studies
of this problem include Frachot \textit{et al}. \cite{FrMoRo04}, Bee
\cite{Bee05}, Chernobai \textit{et al}. \cite{ChMeTrRa05}, Mignola
and Ugoccioni \cite{MiUg06}, Luo \textit{et al}. \cite{LuShDo07},
and Baud \textit{et al}. \cite{BaFrRo03}. A consistent procedure to
adjust for missing data is to fit the data above the threshold using
the correct conditional density. To demonstrate, consider one risk
cell only, where the loss events follow a Poisson process, so that
the annual counts $N(t)$, $t = 1,\ldots,T$ are independent and
Poisson distributed, $Poisson(\lambda )$, with the probability
function

\begin{equation}
\label{Poisson_pdf} p(k|\lambda) = \Pr [N(t) = k] = \frac{\lambda
^k}{k!}\exp ( - \lambda ),\quad \lambda > 0,\;k = 0,1,\ldots\quad .
\end{equation}

\noindent Assume that the severities $X^{(i)}(t)$ are all
independent and identically distributed (iid) from the density
$f(x\vert {{\bm \beta }})$ whose distribution is denoted $F(x\vert
{{\bm \beta }})$, where ${{\bm \beta }}$ is a vector of distribution
parameters. Also, assume that the counts and severities are
independent. Then the loss events above the level $L$ have iid
counts $\tilde {N}(t)$ from $Poisson(\lambda _L )$, $\lambda _L =
\lambda (1 - F(L\vert {{\bm \beta }}))$ and iid severities $\tilde
{X}^{(i)}(t)$ from the conditional density

\begin{equation}
\label{TruncDensity_eq} f_L (x\vert {{\bm \beta }}) = \frac{f(x\vert
{{\bm \beta }})}{1 - F(L\vert {{\bm \beta }})};\quad L \le x <
\infty .
\end{equation}

The joint density (likelihood) of the data ${\rm {\bf Y}}$ over a
period of $T$ years (all counts $\tilde {N}(t)$ and severities
$\tilde {X}^{(i)}(t)$, $i = 1,\ldots,\tilde {N}(t)$, $t =
1,\ldots,T)$, at $\tilde {N}(t)=\tilde {n}(t)$ and $\tilde
{X}^{(i)}(t)=\tilde {x}^{(i)}(t)$, is

\begin{equation}
\label{likelihoodTruncData_eq} l({\rm {\bf y}}\vert {{\bm \beta
}},\lambda) = \prod\limits_{t = 1}^T {p\left( {\tilde {n}(t)\vert
\lambda (1 - F(L\vert {{\bm \beta }}))} \right)\prod\limits_{i =
1}^{\tilde {n}(t)} {f_L\left({\tilde {x}^{(i)}(t)\vert {{\bm \beta
}}} \right)} }.
\end{equation}

\noindent Note that here, the conditional density $f_L (x\vert {{\bm
\beta }})$, rather than $f(x\vert {{\bm \beta }})$, is used. The
parameters $({{\bm \beta }},\lambda )$ can be estimated, for
example, by maximizing the likelihood (\ref{likelihoodTruncData_eq})
and their uncertainties can be estimated using the second
derivatives of the log-likelihood; see Section
\ref{FreqApproach_subsec}. Then estimated frequency $Poisson(\lambda
)$ and severity $f(x\vert {{\bm \beta }})$ densities are used for
the annual loss calculations.

In the case of constant threshold, the maximum likelihood estimators
(MLEs) for parameters ${{\hat {\bm\beta }}}$ and $\hat {\lambda }$
can be calculated marginally, i.e. ${{\hat {\bm\beta }}}$ is
calculated by maximizing the likelihood of the severities; $\hat
{\lambda }_L $ is calculated using the average of the observed
counts; and finally $\hat {\lambda } = \hat {\lambda }_L / (1 -
F(L\vert {{\hat{\bm\beta}}}))$. However, calculation of their
uncertainties will require the use of the full joint likelihood
(\ref{likelihoodTruncData_eq}). If the observed losses are scaled
before fitting or the reporting threshold has changed over time then
one should consider a model with the threshold varying in time
studied in \cite{ShTe09}. In this case the joint estimation of the
frequency and severity parameters using full likelihood of the data
is required even for parameter point estimators.

Of course, the assumption in the above approach is that missing losses and
reported losses are realizations from the same distribution. Thus the method
should be used with extreme caution if a large proportion of data is
missing.

~

\begin{figure}[t]
\centerline{\includegraphics{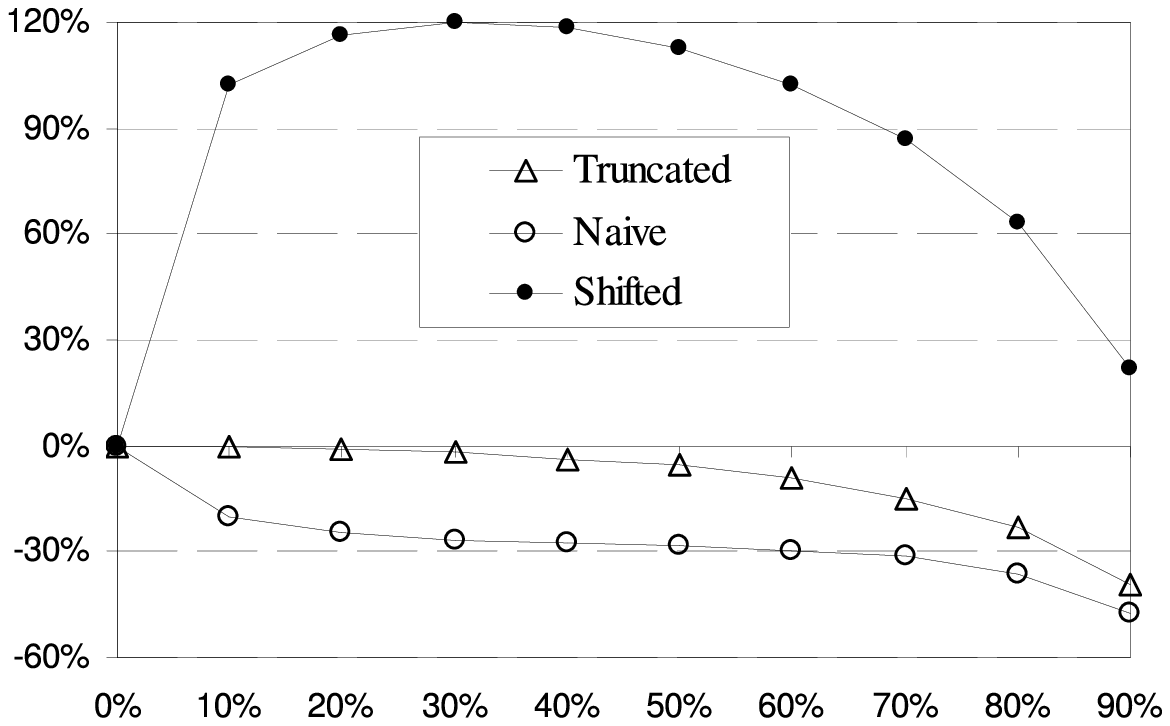}} \caption{Relative bias in
the 0.999 quantile of the annual loss vs {\%} of truncated points
for several models ignoring truncation in the case of light tail
severities from $LN(3,1)$. The annual counts above the truncation
level are from $Poisson(10)$.} \label{TruncDataLightTail_fig}
\end{figure}

\begin{figure}[b]
\centerline{\includegraphics{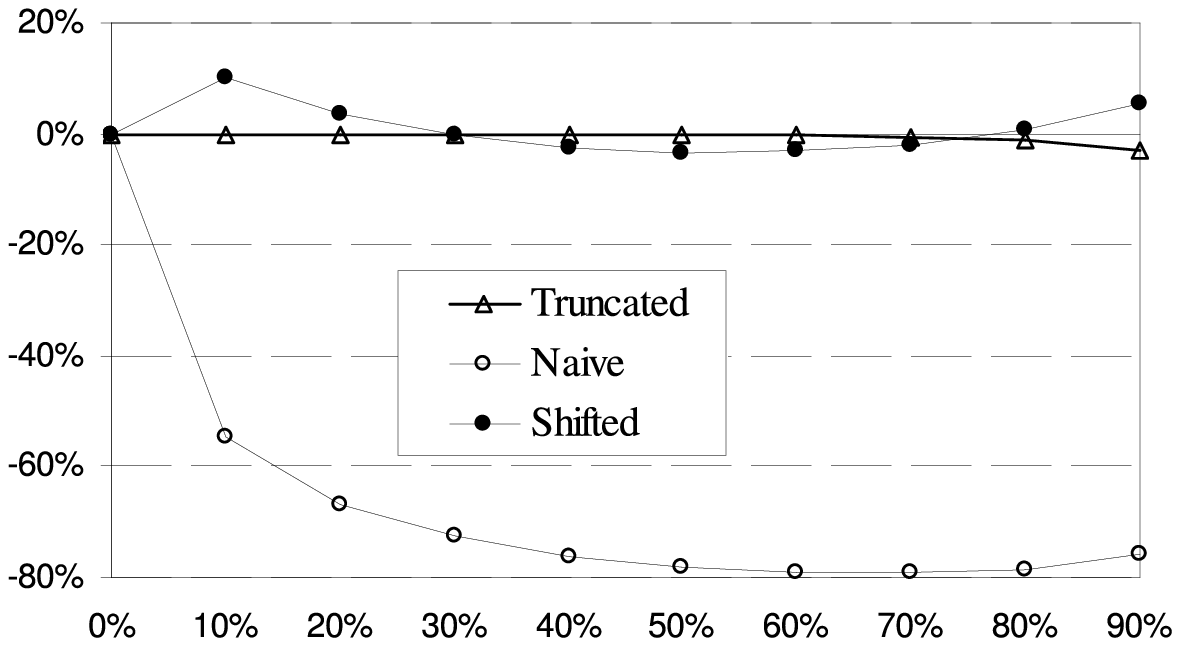}} \caption{Relative bias in
the 0.999 quantile of the annual loss vs {\%} of truncated points
for several models ignoring truncation in the case of heavier tail
severities from $LN(3,2)$. The annual counts above the truncation
level are from $Poisson(10)$.} \label{TruncDataHeavyTail_fig}
\end{figure}

\noindent \textbf{Ignoring missing data.} Ignoring missing data will
have an impact on risk estimates. For example, using data reported
above the threshold, one can fit $Poisson(\lambda _L )$ frequency
and fit the severity using:
\begin{itemize}
\item ``\textit{naive model}'' -- $f(x\vert {{\bm \beta }})$;
\item ``\textit{shifted model}'' -- $f(x - L\vert {{\bm \beta
}})$;
\item ``\textit{truncated model}'' -- $f_L (x\vert {{\bm \beta
}})$.
\end{itemize}
\noindent Calculation of the annual loss quantile using incorrect
frequency and severity distributions will induce a bias. Figure
\ref{TruncDataLightTail_fig} and Figure \ref{TruncDataHeavyTail_fig}
show the relative bias in the 0.999 annual loss quantile (relative
difference between the 0.999 quantiles under the false and true
models) vs a fraction of truncated points for the cases of light and
heavy tail severities respectively. In this example, the severity is
from Lognormal distribution, $LN(\mu ,\sigma )$, i.e. log-severity
$\ln X^{(i)}(t)$ is from the Normal distribution, $Normal(\mu
,\sigma )$, with mean $\mu $ and standard deviation $\sigma $. The
parameter values are chosen the same as some cases considered in
\cite{LuShDo07}. In particular: Figure \ref{TruncDataLightTail_fig}
is the case of $\lambda _L = 10$ and $\sigma = 1$; and Figure
\ref{TruncDataHeavyTail_fig} is the case of $\lambda _L = 10$ and
$\sigma = 2$. The latter corresponds to the heavier tail severity.
Here, the calculated bias is due to the model error only, i.e.
corresponds to the case of a very large data sample. Also note, that
the actual value of the scale parameter $\mu $ is not relevant
because only relative quantities are calculated. ``\textit{Naive
model}'' and ``\textit{shifted model}'' are easy to fit but induced
bias can be very large. Typically: ``\textit{naive model}'' leads to
a significant underestimation of the capital, even for a heavy tail
severity; ``\textit{shifted model}'' is better than ``\textit{naive
model}'' but worse then ``\textit{truncated model}''; the bias from
``\textit{truncated model}'' is less for heavier tail severities.

\section{Modeling severity tail}
\label{sevTail_sec} Due to simple fitting procedure, one of the
popular distributions to model severity is Lognormal. It is a heavy
tail distribution, i.e. belongs to the class of so-called
sub-exponential distributions where the tail decays slower than any
exponential tail. Often, it provides a reasonable overall
statistical fit as reported in the literature, see e.g.
\cite{DuPe06}. Also, it was suggested for OR at the beginning of
Basel II development, see \cite[p.34]{BCBS2001}. However, due to the
high quantile level requirement for OR capital charge, accurate
modeling of extremely high losses (the tail of severity
distribution) is critical and other heavy tail distributions are
often considered to be more appropriate. Two studies of OR data
collected over many institutions are of central importance here:
Moscadelli \cite{Moscadelli04} analysing 2002 LDCE (where Extreme
Value Theory (EVT) is used for analysis in addition to some standard
two parameter distributions) and Dutta and Perry \cite{DuPe06}
analysing 2004 LDCE. The latter paper considered the four-parameter
g-and-h and GB2 distributions as well as EVT and several two
parameter distributions.

~

\noindent \textbf{EVT--threshold exceedances.} There are two types
of EVT models: traditional \textit{block maxima} (modeling the
largest observation) and \textit{threshold exceedances}. The latter
is often used to model the tail of OR severity distribution and is
briefly described below; for more details see McNeil \textit{et al}.
\cite{McFrEm05} and Embrechts \textit{et al}. \cite{EmKlMi97}.
Consider a rv $X$, whose distribution is $\Pr [X \le x] = F(x)$.
Given a threshold $u$, the exceedance of $X$ over $u$ is distributed
from

\begin{equation}
\label{TruncData_df_eq} F_u (y) = \Pr [X - u \le y\vert X > u] =
\frac{F(y + u) - F(u)}{1 - F(u)}.
\end{equation}

\noindent Under quite general conditions, as the threshold $u$
increases, the excess distribution $F_u (.)$ converges to a
generalized Pareto distribution (GPD)

\begin{equation}
\label{GPD_eq} G_{\xi ,\beta } (y) = \left\{ {{\begin{array}{*{20}c}
 {1 - (1 + \xi y / \beta )^{ - 1 / \xi };\;\;\xi \ne 0,} \hfill \\
 {1 - \exp ( - y / \beta );\;\quad \xi = 0.} \hfill \\
\end{array} }} \right.
\end{equation}

\noindent That is we can find a function $\beta (u)$ such that

$$
\mathop {\lim }\limits_{u \to a} \;\mathop {\sup }\limits_{0 \le y
\le a - u} \vert F_u (y) - G_{\xi ,\beta (u)} (y)\vert = 0,
$$

\noindent where $a \le \infty $ is the right endpoint of $F(x)$,
$\xi $ is the shape parameter and $\beta > 0$ is the scale
parameter. Also, $y \ge 0$ when $\xi \ge 0$ and $0 \le y \le - \beta
/ \xi $ when $\xi < 0$. The GPD case $\xi = 0$ corresponds to an
exponential distribution. If $\xi > 0$, the GPD is heavy tailed and
some moments do not exist. In particular, if $\xi \ge 1 / m$ then
the $m$-th and higher moments do not exist. The analysis of OR data
in \cite{Moscadelli04} reported even the cases of $\xi \ge 1$ for
some business lines, i.e. infinite mean distributions; also see
discussions in Ne\v{s}lehov\'a \textit{et al}. \cite{NeEmCh06}. It
seems that the case of $\xi < 0$ is not relevant to modeling OR as
all reported results indicate non-negative shape parameter. Though,
one could think of a risk control mechanism restricting the losses
by an upper level and then the case of $\xi < 0$ might be relevant.

In the context of OR, given iid losses $X^{(k)}$, $k = 1,2,\ldots,K$
one can chose a threshold $u$ and model the losses above the
threshold using GPD (\ref{GPD_eq}) and the losses below using
empirical distribution, i.e.

\begin{equation}
\label{GPD_Emperical_mixture_eq} F(x) \approx \left\{
{{\begin{array}{c}
 {G_{\xi ,\beta } (x - u)(1 - F_n (u)) + F_n (u);\;\;x \ge u,}\\
 {F_n (x) = \frac{1}{K}\sum\nolimits_{k = 1}^K
{I(X^{(k)} \le x)}; \quad x < u.}
\end{array} }} \right.
\end{equation}

\noindent Here, $I(.)$ is an indicator function. There are various
ways to fit the GPD parameters including the Bayesian inference and
maximum likelihood methods; see Section \ref{ModelEstimation_sec}
and Embrechts \textit{et al}. \cite[Section 6.5]{EmKlMi97}. The
approach (\ref{GPD_Emperical_mixture_eq}) is a so-called splicing
method when the density is modeled as

\begin{equation}
\label{mixture_pdf_eq} f(x) = w_1 f_1 (x) + w_2 f_2 (x), \quad w_1 +
w_2 = 1,
\end{equation}

\noindent where $f_1 (x)$ and $f_2 (x)$ are proper density functions
defined on $x < L$ and $x \ge L$ respectively. In
(\ref{GPD_Emperical_mixture_eq}), $f_1 (x)$ is modeled by the
empirical distribution but one may choose a parametric distribution
instead. Splicing can be viewed as a mixture of distributions
defined on non-overlapping regions while a standard mixture
distribution is a combining of distributions defined on the same
range. The choice of the threshold $u$ is critical, for details of
the methods to choose a threshold we refer to \cite[Section
6.5]{EmKlMi97}.

~

\noindent\textbf{\textit{g}-and-\textit{h}, GB2 and GCD
distributions.} A rv $X$ is said to have g-and-h distribution if

\begin{equation}
\label{gandh_transform_eq} X = a + b\frac{\exp (gY) - 1}{g}\exp
(hY^2 / 2),
\end{equation}

\noindent where $Y$ is a rv from the standard Normal distribution
and $(a,b,g,h)$ are the parameters. A comparison of the g-and-h with
EVT was studied in Degen \textit{et al}. \cite{DeEmLa07}. It was
demonstrated that for the g-and-h distribution, convergence of the
excess distribution to the GPD is extremely slow. Therefore,
quantile estimation using EVT may lead to inaccurate results if data
are well modeled by the g-and-h distribution.

GB2 (the generalized Beta distribution of the second kind) is another
four-parameter distribution that nests many important one- and two-parameter
distributions. Its density is defined as

\begin{equation}
\label{GB2_pdf} f(x) = \frac{\vert a\vert x^{ap - 1}}{b^{ap}B(p,q)[1
+ (x / b)^a]^{p + q}},\quad x > 0,
\end{equation}

\noindent where $B(p,q)$ is the Beta function and ($a$,$b$,$p$,$q)$
are parameters. Both g-and-h and GB2 four parameter distributions
were used in \cite{DuPe06} as the alternative to EVT.

A convenient distribution recently suggested for OR in
\cite{GuNiPrRo06} is Generalized Champernowne distribution (GCD)
with the density defined as
\begin{equation}
f(x)=\frac{\alpha
(x+c)^{\alpha-1}((M+c)^\alpha-c^\alpha)}{((x+c)^\alpha+(M+c)^\alpha-2c^\alpha)^2},\;x\geq
0
\end{equation}
and parameters $\alpha>0$, $M>0$, $c\geq 0$. It behaves as Lognormal
in a middle and as Pareto in the tail.

\section{Calculating compound distribution}
\label{CompDFcalc_sec} If the severity and frequency distributions
and their parameters are known then, in general, the distribution
$H(.)$ of the annual loss

\begin{equation}
\label{CompLossModel}
Z = \sum_{i = 1}^N {X^{(i)}}
\end{equation}
should be calculated numerically. Here, we assume that severities
$X^{(i)}$ are iid from the distribution $F(x)$ and the frequency
$N$, with $p_n=\mathrm{Pr}[N=n]$, is independent from the
severities. Most popular numerical methods are Monte Carlo, Panjer
recursion and FFT methods described below. Also, there are several
approximations discussed below too.

\subsection{Monte Carlo method.}
\label{compDistr_MC_sec}
The easiest to implement is Monte Carlo
method with the following logical steps:

%\noindent\hrulefill

\begin{enumerate}
\item For $k = 1,\ldots,K$

\begin{enumerate}
\item Simulate the annual number of events $N$ from the frequency
distribution.

\item Simulate independent severities $X^{(1)},\ldots,X^{(N)}$
from the severity distribution.

\item Calculate $Z^{(k)} = \sum\nolimits_{i = 1}^N {X^{(i)}}$.
\end{enumerate}

\item Next $k$
\end{enumerate}
%\noindent\hrulefill

\noindent Obtained $Z^{(1)},\ldots,Z^{(K)}$ are samples from a
compound distribution $H(.)$. The 0.999 quantile and other
distribution characteristics can be estimated using the simulated
samples in the usual way. Denote the samples
$Z^{(1)},\ldots,Z^{(K)}$ sorted into ascending order as $\tilde
Z^{(1)} \le \ldots \le \tilde Z^{(K)}$, then the quantile $H^{- 1}
(q)$ can be estimated by $\tilde Z^{(\left\lfloor {Kq + 1}
\right\rfloor )}$. Here, $\left\lfloor . \right\rfloor $ denotes
rounding downward. The numerical error (due to the finite number of
simulations $K$) in the quantile estimator can be assessed by
forming a conservative confidence interval $[\tilde Z^{(r)},\tilde
Z^{(s)}]$ to contain the true value with probability $\gamma $. This
can be done by utilizing the fact that the number of samples not
exceeding the quantile $H^{- 1} (q)$ has a Binomial distribution
with parameters $q$ and $K$ (i.e. with $\mathrm{mean} = Kq$ and
${\mathrm{var}} = Kq(1 - q))$. Approximating the Binomial by the
Normal distribution leads to a simple formula:

\begin{equation}
\label{QuatileBounds_eq}
\begin{array}{l}
 r = \left\lfloor l \right\rfloor {\kern 1pt} {\kern 1pt} ,\;\;l = Kq -
F_N^{ - 1} ((1 + \gamma ) / 2)\sqrt {Kq(1 - q)} , \\
 s = \left\lceil u \right\rceil {\kern 1pt} ,\;\;u = Kq + F_N^{ - 1} ((1 +
\gamma ) / 2)\sqrt {Kq(1 - q)} , \\
 \end{array}
\end{equation}

\noindent where $\left\lceil . \right\rceil $ denotes rounding
upwards and $F_N^{ - 1}$ is the standard Normal distribution. The
above formula works very well for $Kq(1 - q) \gtrsim 50$.

A large number of simulations, typically $K \ge 10^5$, should be
used to get a good numerical accuracy for the 0.999 quantile.
However, a priori, the number of simulations required to achieve a
specific accuracy is not known. One of the approaches is to continue
simulations until a numerical error, calculated using
(\ref{QuatileBounds_eq}), achieves the desired level.

\subsection{FFT and Panjer recursion.} Although Monte Carlo
is straightforward and robust, it is slow to get accurate results.
High precision results are especially important for sensitivity
studies, where the first or even the second order derivatives are
involved. Fast Fourier Transform (FFT) and Panjer recursion are the
other two popular alternatives to calculate the distribution of
compound loss (\ref{CompLossModel}). Both have a long history, but
their applications to computing very high quantiles of the compound
distributions in the case of high frequencies or heavy tail
severities are relatively recent developments in the area of
quantitative risk.

~

\noindent\textbf{Panjer recursion.} The Panjer recursion is based on
calculating the compound distribution via convolutions. Using a well
known fact that the distribution of the sum of two independent
continuous random variables can be calculated as convolution, the
compound distribution of the annual loss $Z$ can be calculated as

\begin{equation}
\label{CompLossConvolution_eq} H(z)=\sum_{k=0}^{\infty}\Pr[Z\leq
z|N=k]\Pr[N=k]=\sum_{k=0}^{\infty}p_k F^{(k)}(z),
\end{equation}

\noindent where $F^{(k)}(z)=\Pr[X_1+\cdots+X_k\leq z]$ is the $k$th
convolution of $F$ calculated recursively as
$$
F^{(k)}(z)=\int_0^{z}F^{(k-1)}(z-x)f(x)dx
$$
\noindent with $F^{(0)}(z)=1$, $z\geq 0$ and $F^{(0)}(z)=0$, $z<0$.
Note that integration limits are $0$ and $z$ because severities
considered are nonnegative. Thought the formula is analytic, its
direct calculation is difficult as the convolution powers are not
available in closed form in general. Panjer recursion is a very
efficient method to calculate (\ref{CompLossConvolution_eq}) in the
case of Poisson, negative binomial and binomial frequency
distributions and discrete severities. More precisely, the frequency
distribution should belong to a Panjer class $(a,b,1)$, i.e. to
satisfy
\begin{equation}
\label{PanjerClass_eq} p_n=\left(a+ b/n\right)p_{n-1}, \quad
\mbox{for } n\geq 2.
\end{equation}

\noindent The continuous severity distribution $F(x)$ can be
discretized on $\{0,\Delta,2\Delta,\ldots\}$, by choosing a unit
$\Delta>0$ and defining discrete density
$f_n=\mathrm{Pr}[X=n\Delta]$ as e.g.

\begin{equation}
\label{SeverityDiscretization_eq}
 f_0=F(\Delta/2), \quad
f_n=F(n\Delta+\Delta/2)-F(n\Delta-\Delta/2),\quad n=1,2,\ldots.
\end{equation}

\noindent Then the discrete compound density
$h_n=\mathrm{Pr}[Z=n\Delta]$ can be calculated recursively as

\begin{equation}
\label{ExtendedPanjerRecursion_eq}
h_n=\frac{(p_1-(a+b)p_0)f_n+\sum_{j=1}^{n}\left(a+bj/n\right)f_j
h_{n-j}}{1-af_0}, \quad n\geq 1
\end{equation}
\noindent with $h_0=\sum\limits_{k = 0}^\infty {(f_0)^k p_k}$. The
corresponding discrete distribution converges to the true continuous
distribution $H(z)$ weakly as $\Delta \rightarrow 0$ and the number
of required operations is of the order of $O(n^2)$ (in comparison
with $O(n^3)$ of explicit convolution). Detailed description of this
method can be found in \cite[Section 6.6]{Panjer06} and generalized
versions of Panjer recursion are discussed in \cite{HeLiSc02},
\cite{GeScWa09}, \cite{Sundt92}.

~

\noindent\textbf{FFT method.} The characteristic function (CF) of
the compound loss $Z$ can be calculated as

\begin{equation}
\label{CompCF_eq} \chi (t) = \sum\limits_{k = 0}^\infty {[\varphi
(t)]^kp_k} = \psi (\varphi (t)),
\end{equation}
where $\varphi(t)$ is the CF of the severity density $f(x)$: $
\varphi(t)=\int\nolimits_{ -\infty }^\infty {f(x)\exp(itx)dx}$ and
$\psi (s)$ is the probability generating function of the frequency
distribution: $\psi (s) = \sum\nolimits_{k = 0}^\infty {s^k p_k}$.
For example, in the case of $N$ distributed from $Poisson(\lambda
)$, $\chi(t) = \exp (\lambda \varphi (t) - \lambda)$. Given CF, the
density of $Z$ can be calculated via the inverse Fourier transform
as $ h(z) = \frac{1}{2\pi }\int\nolimits_{-\infty}^\infty
{\chi(t)\exp(-itz) dt}$.  For severity discretized as
$(f_0,f_1,\ldots,f_{M-1})$, e.g. using
(\ref{SeverityDiscretization_eq}), the continuous Fourier transform
is reduced to the discrete Fourier transformation (DFT)

\begin{equation}
\label{DFT_eq}
 \varphi_k=\sum_{m=0}^{M-1}f_m\exp\left(\frac{2\pi
m}{M}k\right),\quad k=0,1,\ldots,M-1.
\end{equation}

\noindent Then the compound loss discrete CF $\chi_k=\psi (\varphi
_k)$, $k=0,\ldots,M-1$ is calculated and the discrete density $h_k$
is recovered from $\chi_k$ by the inverse transformation

\begin{equation}
\label{DFTinv_eq}
h_n=\frac{1}{M}\sum_{k=0}^{M-1}\chi_k\exp\left(-\frac{2\pi
k}{M}n\right),\quad n=0,1,\ldots,M-1.
\end{equation}
\noindent FFT is a method that allows to calculate the above DFTs
(\ref{DFT_eq}-\ref{DFTinv_eq}) efficiently using $O(M\log_2 M)$
operations, when $M=2^r$ with a positive integer $r$; see e.g.
\cite[Chapter 12]{PrTeVeFl02} for details and code. A commonly
recognised pitfall of FFT in evaluating compound distribution, which
was recently studied in great detail in \cite{EmFr08,ScTe08}, is the
so-called aliasing error. To explain, assume that there is no
truncation error in severity discretisation, then FFT procedure
calculates the compound distribution on $m=0,\ldots,M-1$, i.e. the
mass of compound distribution beyond $M-1$ is ``wrapped" into the
range $m=0,\ldots,M-1$. This error is larger for heavy tailed
severities. Choosing $M$ much larger than the required quantile to
reduce this error is very inefficient and the recommended procedure
is to use \emph{tilting}, the transformation increasing the severity
tail decay that commutes with convolution. The \emph{tilting}
transforms the severity as $\tilde f_j=\exp(-j\theta)f_j, \quad
j=0,\ldots, M-1$. Then  FFT calculation of the compound distribution
(\ref{DFT_eq}-\ref{DFTinv_eq}) is performed using transformed
severity and the final result $\tilde h_j$ is adjusted to get the
compound distribution $h_j=\tilde h_j\exp(\theta j)$. It was
reported that the choice $\theta\approx 20/M$ for standard double
precision (8 bytes) calculations works well.

~

\noindent The Panjer recursion has often been compared with FFT, and
it is accepted that the former is slower if the grid size is large,
see e.g. \cite{EmFr08}. Both methods have discretization error. Note
that, Panjer recursion has no truncation error presented in FFT.
Also note that, FFT can be used in general for any frequency and
severity distributions while Panjer recursion is restricted to
non-negative severities and a special class of frequency
distributions. Typically, both methods are faster than Monte Carlo
by a factor of few orders. Other methods to calculate the compound
distribution include direct integration of the CF \cite{LuSh09} and
a hybrid method combining Panjer recursion, importance sampling and
trans-dimensional Markov chain Monte Carlo considered in
\cite{PeJoDo07}.

\subsection{Closed-form approximation}
The moments of compound distributions can be expressed via the
moments of frequency and severity distributions. This can be
utilized to approximate the compound loss using e.g. Normal or
translated Gamma  distributions by matching two or three moments
respectively, see e.g. \cite[Section 10.2.3]{McFrEm05}. Of course
for heavy tailed distributions and high quantiles these
approximations do not work well (note that low order moments may not
even exist for some ORs). If the severities $X^{(i)}$ are iid from
the sub-exponential (heavy tail) distribution $F(.)$, then the tail
of the compound distribution $H(.)$ is related to the severity tail
as

\begin{equation}
\label{ExtremeTailApprox_eq}
 1 - H(z)\sim E[N](1 - F(z)),\quad z \to \infty ,
\end{equation}

\noindent see  e.g. \cite[Theorem 1.3.9]{EmKlMi97}. Here, ``$\sim
$'' means that the ratio of the left- and right-hand sides converge
to 1. The validity of this asymptotic result was demonstrated for
the cases when $N$ is distributed from Poisson, binomial or negative
binomial. This approximation can be used to calculate the quantiles
of the annual loss as

\begin{equation}
\label{VaRtailApprox_eq} VaR_q (Z)\sim F^{ - 1} \left( {1 - \frac{1
- q}{E[N]}} \right),\quad q \to 1.
\end{equation}

\noindent For application in the OR context, see B\"{o}cker and
Kl\"{u}ppelberg \cite{BoKl05}. Under the assumption that the
severity has finite mean, B\"{o}cker and Sprittulla \cite{BoSp06}
derived a correction reducing the approximation error of
(\ref{VaRtailApprox_eq}).

\section{Model fitting}
\label{ModelEstimation_sec} Estimation of the frequency and severity
distributions is a challenging task, especially for low frequency
high impact losses, due to very limited data for some risks. The
main tasks involved into the fitting are: finding the best point
estimates for the distribution parameters, quantification of the
parameter uncertainties, assessing the model quality (model error).
In general, these tasks can be accomplished by undertaking either
the so-called frequentist or Bayesian approaches briefly discussed
below.

\subsection{Frequentist approach}
\label{FreqApproach_subsec}
Fitting distribution parameters using
data via the frequentist approach is a classical problem described
in many textbooks. For the purposes of this review it is worth to
mention several aspects and methods. Firstly, under the frequentist
approach one says that the model parameters are fixed while their
estimators have associated uncertainties that typically converge to
zero when a sample size increases. Several popular methods to fit
parameters of the assumed distribution are:
\begin{itemize}
\item method of moments -- matching the observed moments;

\item matching certain quantiles of empirical distribution;

\item maximum likelihood -- find parameter values that maximize the
joint likelihood of data;

\item estimating parameters by minimizing a certain distance between
empirical and theoretical distributions, e.g. Anderson-Darling or
other statistics, see Ergashev \cite{Ergashev08}.

\end{itemize}

\noindent The most popular approach is the maximum likelihood
method. Here, given the model parameters ${{\bm \theta }} = (\theta
_1 ,\theta _2 ,\ldots)$, assume that the joint density (likelihood)
of data ${\rm {\bf Y}} = (Y_1 ,\ldots,Y_n )$ is known in functional
form $\ell ({\rm {\bf y}}\vert {{\bm \theta }})$. Then the maximum
likelihood estimators (MLEs) ${{\hat {\bm\theta }}}^{MLE}$ are the
values of the parameters ${{\bm \theta }}$ maximizing $\ell ({\rm
{\bf y}}\vert {{\bm \theta }})$. Often it is assumed that
$Y_1,\ldots,Y_n $ are iid from $f(.\vert {{\bm \theta }})$; then the
likelihood is $\ell ({\rm {\bf y}}\vert {{\bm \theta }}) =
\prod\limits_{i = 1}^n {f(y_i \vert {{\bm \theta }})}$. The
uncertainty of the MLEs can be estimated using the asymptotic result
that: under suitable regularity conditions, as the sample size
increases, ${{\hat {\bm\theta }}}^{MLE}$ converges to ${{\bm \theta
}}$ and is Normally distributed with the mean ${{\bm \theta }}$ and
covariance matrix $n^{ - 1}{\rm {\bf I}}({{\bm \theta }})^{ - 1}$,
where

\begin{equation}
\label{InformationMatrix_eq} {\rm {\bf I}}({{\bm \theta }})_{km} = -
\frac{1}{n}E[\partial ^2\ln \ell({\rm{\bf Y}}\vert {{\bm \theta }})
/
\partial \theta _k
\partial \theta _m ]
\end{equation}

\noindent is the expected Fisher information matrix. For precise
details on regularity conditions and proofs see e.g Lehmann
\cite[Theorem 6.2.1 and 6.2.3]{Lehmann83}. The required regularity
conditions are mild but often difficult to prove. Also, whether a
sample size is large enough to use this asymptotic result is another
difficult question to answer in practice. Often,
(\ref{InformationMatrix_eq}) is approximated by the observed
information matrix

\begin{equation}
\label{ObservedInformationMatrix_eq} -\frac{1}{n}\partial ^2\ln
\ell({\mathrm{\bf y}}\vert {{\bm \theta }}) /
\partial \theta _k
\partial \theta _m = - \frac{1}{n}\sum\nolimits_{i = 1}^n {\partial ^2\ln f(y_i \vert {{\bm
\theta }}) / \partial \theta _k \partial \theta _m }
\end{equation}

\noindent for a given realization of data ${\mathrm{\bf y}}$. This
should converge to the matrix (\ref{InformationMatrix_eq}) by the
law of large numbers. Note that the mean and covariances depend on
the unknown parameters ${{\bm \theta }}$ and are usually estimated
by replacing ${{\bm \theta }}$ with ${{\hat {\bm\theta }}}^{MLE}$
for a given realization of data. This asymptotic approximation may
not be accurate enough for small samples.

Another common way to estimate the parameter uncertainties is
Bootstrap method. It is based on generating many data samples of the
same size from the empirical distribution of the original sample and
calculating the parameter estimates for each sample to get the
distribution of the estimates. For a good introduction to the method
we refer the reader to Efron and Tibshirani \cite{EfTi93}.

Usually maximization of the likelihood (or minimization of some
distances in other methods) should be done numerically. Popular
numerical optimization algorithms include: simplex method, Newton
methods, expectation maximization (EM) algorithm, simulated
annealing. It is worth to mention that the last is attempting to
find a global maximum while other methods are designed to find a
local maximum only. For the latter, using different starting points
helps to find a global maximum. Typically, EM is more stable and
robust than the standard deterministic methods such as simplex or
Newton methods. To assess the quality of the fit, there are several
popular goodness of fit tests including Kolmogorov-Smirnov,
Anderson-Darling and chi-square tests. Also, the likelihood ratio
test and Akaike's information criterion are often used to compare
models. Again, detailed descriptions of the above mentioned
methodologies can be found in many textbooks; for application in OR
context see e.g. Panjer \cite{Panjer06}.

\subsection{Bayesian inference approach}
\label{BayesianApproach_subsec} There is a broad literature covering
Bayesian inference and its applications for the insurance industry
as well as other areas. For a good introduction to the Bayesian
inference method, see Berger \cite{Berger85}. In our opinion, this
approach is well suited for OR, though it is not often used in the
OR literature; it was briefly mentioned in books Cruz \cite{Cruz02}
and Panjer \cite{Panjer06} and was applied to OR modeling in several
recent papers referred to below. To sketch the method, consider a
random vector of data ${\rm {\bf Y}} = (Y_1,\ldots,Y_n )$ whose
density, for a given vector of parameters ${{\bm \theta }} = (\theta
_1 ,\theta _2 ,\ldots)$, is $\ell ({\rm {\bf y}}\vert {{\bm \theta
}})$. In the Bayesian approach, both data and parameters are
considered to be random. A convenient interpretation is to think
that the parameter is a rv with some distribution and the true value
(which is deterministic but unknown) of the parameter is a
realization of this rv. Then Bayes' theorem can be formulated as

\begin{equation}
\label{BayesTheorem_eq} h({\rm {\bf y}},{{\bm \theta }}) = \ell
({\rm {\bf y}}\vert {{\bm \theta }})\pi ({{\bm \theta }}) = \pi ({
{\bm \theta }}\vert {\rm {\bf y}})h({\rm {\bf y}}),
\end{equation}

\noindent where $\pi ({{\bm \theta }})$ is the density of parameters
(a so-called prior density); $\pi ({{\bm \theta }}\vert {\rm {\bf
y}})$ is the density of parameters given data ${\rm {\bf Y}}$ (a
so-called posterior density); $h({\rm {\bf y}},{{\bm \theta }})$ is
the joint density of the data and parameters; $\ell ({\rm {\bf
y}}\vert {{\bm \theta }})$ is the density of data for given
parameters (likelihood); and $h({\rm {\bf y}})$ is a marginal
density of ${\rm {\bf Y}}$. If $\pi ({{\bm \theta }})$ is continuous
then $h({\rm {\bf y}}) = \int {\ell ({\rm {\bf y}}\vert {{\bm \theta
}})\pi ({{\bm \theta }})d{{\bm \theta }}} $ and if $\pi ({{\bm
\theta }})$ is a discrete, then the integration should be replaced
with a corresponding summation. Typically, $\pi ({{\bm \theta }})$
depends on a set of further parameters, the so-called
hyper-parameters, omitted here for simplicity of notation. The
choice and estimation of the prior will be discussed in Section
\ref{EstimatingPrior_subsec}. Using (\ref{BayesTheorem_eq}), the
posterior density can be written as

\begin{equation}
\label{PosteriorDensity_eq} \pi ({{\bm \theta }}\vert {\rm {\bf y}})
= \ell ({\rm {\bf y}}\vert {{\bm \theta }})\pi ({{\bm \theta }}) /
h({\rm {\bf y}}).
\end{equation}

\noindent Here, $h({\rm {\bf y}})$ plays the role of a normalization
constant and the posterior can be viewed as a combination of a prior
knowledge contained in $\pi ({{\bm \theta }})$ with the data
likelihood $\ell ({\rm {\bf y}}\vert {{\bm \theta }})$.

If the data $Y_1,\ldots,Y_n $ are conditionally (given ${ {\bm
\theta }})$ iid then the posterior can be calculated iteratively,
i.e. the posterior distribution calculated after $k-1$ observations
can be treated as a prior distribution for the $k$-th observation.
Thus the loss history over many years is not required, making the
model easier to understand and manage, and allowing experts to
adjust the priors at every step.

In practice, it is not unusual to restrict parameters. In this case
the posterior distribution will be a truncated version of the
posterior distribution in the unrestricted case. For example, if we
identified that ${{\bm \theta }}$ is restricted to some range $[{
{\bm \theta }}_L ,{{\bm \theta }}_H ]$ then the posterior will have
the same type as in the unrestricted case but truncated outside this
range.

Sometimes the posterior density can be calculated in closed form.
This is the case for the so called conjugate prior distributions
where the prior and posterior distributions are of the same type,
for a precise definition, see e.g. \cite[Section 4.2.2,
p.130]{Berger85}

The mode, mean or median of the posterior $\pi ({{\bm \theta }}\vert
{\rm {\bf y}})$ are often used as point estimators for the parameter
${{\bm \theta }}$, though in OR context we recommend the use of the
whole posterior as discussed in Section \ref{CapitalViaPredDF_sec}.
Popular model selection criteria include the Deviance Information
Criterion (DIC) and Bayes Information Criterion (BIC); see e.g.
Peters and Sisson \cite{PeSi06} in the OR context.

~

\noindent{\textbf{Gaussian approximation.}} A Gaussian approximation
for the posterior $\pi ({{\bm \theta }}\vert {\rm {\bf y}})$ is
obtained by a second order Taylor series expansion around the mode
${{\hat {\bm\theta }}}$

\begin{equation}
\label{PosteriorGaussianApprox_eq} \ln \pi ({{\bm \theta }}\vert
{\rm {\bf y}}) \approx \ln \pi ({{\hat {\bm\theta }}}\vert {\rm {\bf
y}}) + \frac{1}{2}\sum\limits_{i,j} {\left. {\frac{\partial ^2\ln
\pi ({{\bm \theta }}\vert {\rm {\bf y}})}{\partial \theta _i
\partial \theta _j }} \right|_{{{\bm \theta }} = {{\hat
{\bm\theta }}}} (\theta _i - \hat {\theta }_i )(\theta _j - \hat
{\theta }_j )} ,
\end{equation}

\noindent if the prior is continuous at ${{\hat {\bm\theta }}}$.
Under this approximation, $\pi ({{\bm \theta }}\vert {\rm {\bf y}})$
is a multivariate Normal distribution with the mean ${{\hat
{\bm\theta }}}$ and covariance matrix ${{\bm \Sigma }} = {\rm{\bf{
I}}}^{ - 1}$,

$$({\rm {\bf I}})_{ij} = \left. { - \partial ^2\ln \pi
({{\bm \theta }}\vert {\rm {\bf y}}) / \partial \theta _i
\partial \theta _j } \right|_{{{\bm \theta }} = {{\hat
{\bm\theta }}}}.$$

\noindent In the case of improper constant priors, this
approximation compares to the Gaussian approximation for the MLEs
(\ref{ObservedInformationMatrix_eq}). Also, note that in the case of
constant priors, the mode of the posterior and MLE are the same.
This is also true if the prior is uniform within a bounded region,
provided that the MLE is within this region.

~

\noindent \textbf{Markov chain Monte Carlo methods.} In general,
estimation (sampling) of the posterior numerically can be
accomplished using Markov chain Monte Carlo (MCMC) methods, see e.g.
Robert and Casella \cite[Sections 6-10]{RoCa04} for widely used
Metropolis-Hastings and Gibbs sampler algorithms. In particular,
\textit{Random Walk Metropolis-Hastings} (RW-MH) \textit{within
Gibbs} algorithm is easy to implement and often efficient if the
likelihood function can be easily evaluated. It is referred to as
\textit{single-component Metropolis-Hastings} in Gilks \textit{et
al}. \cite[Section 1.4]{Gilks96}. The algorithm is not well known
among OR practitioners and we would like to mention its main
features; see e.g. \cite{ShTe09} for application in the context of
OR and Peters \textit{et al}. \cite{PeShWu09} for application in the
context of a similar problem in the insurance. The RW-MH within
Gibbs algorithm creates a reversible Markov chain with a stationary
distribution corresponding to our target posterior distribution.
Denote by ${{\bm \theta }}^{(m)}$ the state of the chain at
iteration $m$. The algorithm proceeds by proposing to move the $i$th
parameter from the current state $\theta _i ^{(m - 1)}$ to a new
proposed state $\theta _i^\ast $ sampled from the MCMC proposal
transition kernel. Then the proposed move is accepted according to
some rejection rule derived from a reversibility condition. Note
that, here the parameters are updated one by one while in a general
Metropolis-Hastings algorithm all parameters are updated
simultaneously. Typically, the parameters are restricted by simple
ranges, $\theta_i \in [a_i ,b_i ]$, and proposals are sampled from
Normal distribution. Then the logical steps of the algorithm are as
follows:

%\noindent\hrulefill
%\noindent\textbf{RW-MH within Gibbs algorithm}
\begin{enumerate}
\item Initialize $\theta _i^{(m = 0)} $, $i = 1,\ldots,I$ by e.g. using
MLEs.

\item For $m = 1,\ldots,M$

\begin{enumerate}
\item Set ${{\bm \theta }}^{(m)} = {{\bm \theta }}^{(m - 1)}$

\item For $i = 1,\ldots,I$

\item Sample proposal $\theta _i^\ast $ from the transition kernel,
e.g. from the truncated Normal density
\begin{equation}
\label{TransitionKernel}
f_N^{(T)} (\theta _i^\ast \vert \theta
_i^{(m)} ,\sigma _i ) = \frac{f_N (\theta _i^\ast \vert \theta
_i^{(m)} ,\sigma _i )}{F_N (b_i \vert \theta _i^{(m)} ,\sigma _i ) -
F_N (a_i \vert \theta _i^{(m)} ,\sigma _i )},
\end{equation}
where $f_N (x\vert \mu ,\sigma )$ and $F_N (x\vert \mu ,\sigma )$
are the Normal density and its distribution with mean $\mu $ and
standard deviation $\sigma $.

\item  Accept proposal with the acceptence probability

$$
p({{\bm \theta }}^{(m)},{{\bm \theta }}^\ast ) = \min \left\{
{1,\frac{\pi ({{\bm \theta }}^\ast \vert {\rm {\bf y}})f_N^{(T)}
(\theta _i^{(m)} \vert \theta _i^\ast ,\sigma _i )}{\pi ({{\bm
\theta }}^{(m)}\vert {\rm {\bf y}})f_N^{(T)} (\theta _i^\ast \vert
\theta _i^{(m)} ,\sigma _i )}} \right\},
$$

where ${{\bm \theta }}^\ast = (\theta _1^{(m)} ,\ldots,\theta _{i -
1}^{(m)} ,\theta _i^\ast ,\theta _i^{(m - 1)} ,\ldots)$, i.e.
simulate $U$ from the uniform (0,1) and set $\theta _i^{(m)} =
\theta _i^\ast $ if $U < p({{\bm \theta }}^{(m)},{{\bm \theta
}}^\ast )$. Note that, the normalization constant of the posterior
(\ref{PosteriorDensity_eq}) does not contribute here.

\item Next $i$
\end{enumerate}

\item Next $m$
\end{enumerate}

%\noindent\hrulefill

\noindent This procedure builds a set of correlated samples from the
target posterior distribution. One of the most useful asymptotic
properties is the convergence of ergodic averages constructed using
the Markov chain samples to the averages obtained under the
posterior distribution. The chain has to be run until it has
sufficiently converged to the stationary distribution (posterior
distribution) and then one obtains samples from the posterior
distribution. The RW-MH algorithm is simple in nature and easy to
implement. General properties of this algorithm can be found in e.g.
\cite[Section 7.5]{RoCa04}. However, for a bad choice of the
proposal distribution, the algorithm gives a very slow convergence
to the stationary distribution. There have been several recent
studies regarding the optimal scaling of the proposal distributions
to ensure optimal convergence rates, see e.g. Bedard and Rosenthal
\cite{BeRo08}. The suggested asymptotic acceptance rate optimizing
the efficiency of the process is 0.234. Usually it is recommended
that the $\sigma_i $ in (\ref{TransitionKernel}) are chosen to
ensure that the acceptance probability is roughly close to $0.234$
(this requires some tuning of the $\sigma_i $ prior to final
simulations).

\section{Combining different data sources}
\label{DataCombining_sec} Basel II AMA requires (see
\cite[p.152]{BaselII06}) that: ``\textit{Any operational risk
measurement system must have certain key features to meet the
supervisory soundness standard set out in this section. These
elements must include the use of internal data, relevant external
data, scenario analysis and factors reflecting the business
environment and internal control systems}''.

Combining these different data sources for model estimation is
certainly one of the main challenges in OR. As it was emphasized in
the interview with several industry's top risk executives
\cite{Davis06}: ``\textit{[. . .] Another big challenge for us is
how to mix the internal data with external data; this is something
that is still a big problem because I don't think anybody has a
solution for that at the moment'' }and\textit{ ``What can we do when
we don't have enough data [. . .] How do I use a small amount of
data when I can have external data with scenario generation? [. . .]
I think it is one of the big challenges for operational risk
managers at the moment}''.

Often in practice, accounting for factors reflecting the business
environment and internal control systems is achieved via scaling of data.
Then ad-hoc procedures are used to combine internal data, external data and
expert opinions. For example:
\begin{itemize}
\item Fit the severity distribution to the combined samples of
internal and external data and fit the frequency distribution using
internal data only.

\item Estimate the Poisson annual intensity for the frequency
distribution as $w\lambda _{int} + (1 - w)\lambda _{ext} $, where
the intensities $\lambda _{ext} $ and $\lambda _{int} $ are implied
by the external and internal data respectively, using expert
specified weight $w$.

\item Estimate the severity distribution as a mixture $w_1 F_{SA} (x)
+ w_2 F_I (x) + (1 - w_1 - w_2 )F_E (x)$, where $F_{SA} (x)$, $F_I
(x)$ and $F_E (x)$ are the distributions identified by scenario
analysis, internal data and external data respectively, using expert
specified weights $w_1 $ and $w_2 $.

\item Minimum variance principle -- the combined estimator is a linear
combination of the individual estimators obtained from internal
data, external data and expert opinion separately with the weights
chosen to minimise the variance of the combined estimator.
\end{itemize}

\noindent Probably the easiest to use and flexible procedure is
minimum variance principle. The rationale behind the principle is as
follows. Consider two unbiased independent estimates $\hat {\theta
}^{(1)}$ and $\hat {\theta }^{(2)}$ for parameter $\theta$, i.e.
$E[\hat {\theta }^{(k)}] = \theta $ and ${\rm{var}}(\hat {\theta
}^{(k)})= \sigma_k^2 $, $k = 1,2$. Then the combined unbiased linear
estimator and its variance are

\begin{equation}
\label{MinVarPrincipleTwoSources_eq}
\begin{array}{l}
 \hat {\theta }_{tot} = w_1 \hat {\theta }^{(1)} + w_2 \hat {\theta
}^{(2)},\quad w_1 + w_2 = 1;\quad \\
 {\rm{var}}(\hat {\theta}_{tot} ) = w_1^2 \sigma_1^2 + (1 - w_1 )^2\sigma_2^2 .
\\
 \end{array}
\end{equation}

\noindent It is easy to find that the weights

$$w_1 = \frac{\sigma_2^2 }{\sigma _1^2 + \sigma _2^2},\quad
w_2 = \frac{\sigma_1^2 }{\sigma _1^2 + \sigma _2^2 }$$

\noindent minimize $\rm{var}(\hat {\theta}_{tot} )$. These weights
behave as it is expected in practice. In particular, $w_1 \to 1$ if
$\sigma _1^2 / \sigma _2^2 \to 0$ ($\sigma _1^2 / \sigma _2^2 $ is
the uncertainty of the estimator $\hat {\theta }^{(1)}$ over the
uncertainty of $\hat {\theta }^{(2)})$ and $w_1 \to 0$ if $\sigma
_2^2 / \sigma _1^2 \to 0$. This method can easily be extended to
combine three or more estimators:

\begin{equation}
\label{MinVarPrincipleManySources_eq} \hat {\theta }_{tot} = w_1
\hat {\theta }^{(1)} + \ldots + w_K \hat {\theta }^{(K)},\;\;w_1 +
\ldots + w_K = 1;
\end{equation}

\noindent with
\begin{equation}
w_i=\frac{1/\sigma^2_i}{\sum_{k=1}^{K}(1/\sigma_k^2)}, \quad
i=1,\ldots,K
\end{equation}

\noindent minimizing ${\rm{var}}(\hat {\theta }_{tot} )$.
Heuristically, it can be applied to almost any quantity e.g.
distribution parameter or distribution characteristic such as mean,
variance, etc. The assumption that the estimators are unbiased
estimators for $\theta $ is probably reasonable when combining
estimators from different experts (or from expert and internal
data). However, it is certainly questionable if applied to combine
estimators from the external and internal data. Below, we focus on
the Bayesian inference method that can be used to combine these data
sources in a consistent statistical framework.

\subsection{Bayesian Inference to combine two data sources}
\label{CombiningTwoSources_subsec}
Bayesian inference is a
statistical technique well suited to combine different data sources
for data analysis; for application in OR context, see Shevchenko and
W\"{u}thrich \cite{ShWu06}. For the closely related methods of
credibility theory, see B\"{u}hlmann and Gisler \cite{BuGi05} and
B\"{u}hlmann \textit{et al}. \cite{BuShWu07}.

As in Section \ref{BayesianApproach_subsec}, consider a random
vector of data ${\rm {\bf Y}} = (Y_1,\ldots,Y_n )$ whose density,
for a given vector of parameters ${{\bm \theta }} = (\theta _1
,\ldots,\theta _I )$, is $\ell ({\rm {\bf y}}\vert {{\bm \theta
}})$. Then the posterior distribution (\ref{PosteriorDensity_eq}) is

\begin{equation}
\label{PosteriorWihtoutNormalization_eq} \pi ({{\bm \theta }}\vert
{\rm {\bf y}}) \propto \ell ({\rm {\bf y}}\vert {{\bm \theta }})\pi
({{\bm \theta }}).
\end{equation}

\noindent Hereafter, $ \propto $ is used for statements with the
relevant terms only. The prior distribution $\pi ({{\bm \theta }})$
can be estimated using appropriate expert opinions or using external
data. Thus the posterior distribution $\pi ({{\bm \theta }}\vert
{\rm {\bf y}})$ combines the prior knowledge (expert opinions or
external data) with the observed data using formula
(\ref{PosteriorWihtoutNormalization_eq}). In practice, we start with
the prior $\pi ({{\bm \theta }})$ identified by expert opinions or
external data. Then, the posterior $\pi ({{\bm \theta }}\vert {\rm
{\bf y}})$ is calculated using
(\ref{PosteriorWihtoutNormalization_eq}) when actual data are
observed. If there is a reason (e.g. a new control policy introduced
in a bank), then this posterior can be adjusted by an expert and
treated as the prior for subsequent observations. Examples are
presented in \cite{ShWu06}.

As an illustrative example, consider modeling of the annual counts
using Poisson distribution. Suppose that, given \textit{$\lambda $},
data ${\rm {\bf N}} = (N(1),\ldots,N(T))$ are iid from
$Poisson(\lambda )$ and prior for $\lambda $ is $Gamma(\alpha ,\beta
)$ with a density
$$\pi (\lambda ) = (\lambda / \beta )^{\alpha -
1}\exp ( - \lambda / \beta ) / (\Gamma (\alpha )\beta ),$$

\noindent where $\Gamma (\alpha )$ is a gamma function. Substituting
the prior density and the likelihood of the data $\ell ({\rm {\bf
n}}\vert \lambda ) = \prod\limits_{t = 1}^T {e^{ - \lambda }\lambda
^{n(t)} / n(t)!} $ into (\ref{PosteriorWihtoutNormalization_eq}), it
is easy to find that the posterior is $Gamma(\tilde {\alpha },\tilde
{\beta })$ with parameters $\tilde{\alpha } = \alpha +
\sum\nolimits_{t = 1}^T {n(t)} $ and $\tilde {\beta } = \beta / (1 +
\beta T)$. The expected number of events, given past observations,
(which is a mean of the posterior in this case) allows for a good
interpretation as follows:

\begin{equation}
\label{PosteriorMeanPoisson_eq} E[N(T + 1)\vert {\rm {\bf N}}] =
E[\lambda \vert {\rm {\bf N}}] = \tilde {\alpha }\tilde {\beta } =
w\bar {N} + (1 - w)\lambda _0 ,
\end{equation}

\noindent where $\bar {N} = \frac{1}{T}\sum\nolimits_{t = 1}^T
{n(t)} $ is the MLE of $\lambda $ using the observed counts only;
$\lambda _0 = \alpha \beta $ is the estimate of $\lambda $ using a
prior distribution only (e.g. specified by expert or from external
data); $w = T / (T + 1 / \beta )$ is the credibility weight in [0,1)
used to combine $\lambda _0 $ and $\bar {N}$. As the number of years
$T$ increases, the credibility weight $w$ increases and vice versa.
That is, the more observations we have, the greater credibility
weight we assign to the estimator based on the observed counts,
while the lesser weight is attached to the prior estimate. Also, the
larger the volatility of the prior (larger $\beta )$, the greater
the credibility weight assigned to observations.

One of the features of the Bayesian method is that the variance of
the posterior $\pi ({{\bm \theta }}\vert {{\bm y}})$ converges to
zero for a large number of observations. This means that the true
value of the risk profile ${{\bm \theta }}$ will be known exactly.
However, there are many factors (for example, political, economical,
legal, etc.) changing in time that will not allow precise knowledge
of the risk profile. One can model this by allowing parameters to be
truly stochastic variables as discussed in Section
\ref{Dependence_sec}. Also, the variance of the posterior
distribution can be limited by some lower levels (e.g. 5{\%}) as has
been done in solvency approaches for the insurance industry, see
e.g. Swiss Solvency Test \cite[formulas (25)-(26)]{SST06}.

\subsection{Estimating priors}
\label{EstimatingPrior_subsec}

In general, the structural parameters of the prior distributions can be
estimated subjectively using expert opinions (pure Bayesian approach) or
using data (empirical Bayesian approach).

~

\noindent \textbf{Pure Bayesian approach}. In a pure Bayesian
approach, the prior is specified subjectively (i.e. using expert
opinions). Berger \cite[Section 3.2]{Berger85} lists several
methods:
\begin{itemize}
\item Histogram approach: split the space of ${{\bm \theta }}$
into intervals and specify the subjective probability for each
interval.

\item Relative Likelihood Approach: compare the intuitive likelihoods
of the different values of ${{\bm \theta }}$.

\item CDF determinations: subjectively construct the cumulative
distribution function for the prior and sketch a smooth curve.

\item Matching a Given Functional Form: find the prior distribution
parameters assuming some functional form for the prior to match
beliefs (on the moments, quantiles, etc) as close as possible.
\end{itemize}

\noindent The use of a particular method is determined by the
specific problem and expert experience. Usually, if the expected
values for the quantiles (or mean) and their uncertainties are
estimated by the expert then it is possible to fit the priors; also
see \cite{ShWu06}.

~

\noindent \textbf{Empirical Bayesian approach}. The prior
distribution can be estimated using the marginal distribution of the
observations. The data can be collective industry data, collective
data in the bank, etc. For example, consider a specific risk cell in
$J$ banks with the data ${\rm {\bf Y}}_j =(Y_j(1),\ldots,Y_j (K_j
))$, $j = 1,\ldots,J$. Here, $K_j $ is the number of observations in
bank $j$. Depending on the set up, these could be annual counts or
severities or both. Assume that $Y_j (k)$, $k = 1,\ldots,K_j $ are
iid from $f(.\vert {{\bm \theta }}_j )$, for given ${{\bm \theta
}}_j $, and are independent between different banks; and ${{\bm
\theta }}_j $, $j = 1,\ldots,J$ are iid from the prior $\pi (.)$.
That is, the risk cell in the $j$-th bank has its own risk profile
${{\bm \theta }}_j $, but ${{\bm \theta }}_1 ,\ldots,{{\bm \theta
}}_J $ are drawn from the same distribution $\pi (.)$. One can say
that the risk cells in different banks are the same a priori. Then
the likelihood of all observations can be written as

\begin{equation}
\label{TotalLikelihood_eq} h({\rm {\bf y}}_1 ,\ldots,{\rm {\bf y}}_J
) = \prod\limits_{j = 1}^J {\int {\left[ {\prod\limits_{k = 1}^{K_j
} {f(y_j (k)\vert {{\bm \theta }}_j )} } \right]\pi ({{\bm \theta
}}_j )d{{\bm \theta }}_j } } .
\end{equation}

\noindent Now, the parameters of $\pi ({{\bm \theta }}_j )$ can be
estimated by maximizing the above likelihood. The distribution $\pi
({{\bm \theta }}_j )$ is a prior distribution for the cell in the
$j$-th bank. Then, using internal data of the risk cell in the
$j$-th bank, the posterior $\pi ({{\bm \theta }}_j \vert {\rm {\bf
y}}_j )$ is calculated using
(\ref{PosteriorWihtoutNormalization_eq}).

It is not difficult to include a priori known differences (e.g.
exposure indicators, expert opinions on the differences, etc)
between the risk cells from the different banks. As an example,
consider the case when the annual frequency of the events in the
$j$th bank is modeled by a Poisson distribution with a Gamma prior
and observations $N_j (k)$, $k = 1,\ldots,K_j ,\;j = 1,\ldots,J$.
Assume that, for given $\lambda _j $, $N_j (1),\ldots,N_j (K_j )$
are independent and $N_j (k)$ is distributed from $Poisson(\lambda
_j V{ }_j(k))$. Here, $V_j (k)$ is a known constant (i.e. the gross
income or the volume or combination of several exposure indicators)
and $\lambda _j $ is the risk profile of the cell in the $j$-th
bank. Assuming further that $\lambda_1,\ldots,\lambda_J$ are iid
from a common prior distribution $Gamma(\alpha ,\beta )$, the
likelihood of all observations can be written similar to
(\ref{TotalLikelihood_eq}) and parameters $(\alpha ,\beta )$ can be
estimated using the maximum likelihood or method of moments; see
\cite{ShWu06}. Often it is easier to scale the actual observations
that can be incorporated into the model set up as follows. Given
data $X_j (k)$, $k = 1,\ldots,K_j $, $j = 1,\ldots,J$ (these could
be frequencies or severities), consider variables $Y_j (k) = X_j (k)
/ V{ }_j(k)$. Assume that, for given ${ {\bm \theta }}_j , \quad Y_j
(k)$, $k = 1,\ldots,K_j $ are iid from $f(.\vert {{\bm \theta }}_j
)$ and ${{\bm \theta }}_1 ,\ldots,{{\bm \theta }}_J $ are iid from
the prior $\pi (.)$. Then again one can construct the likelihood of
all data similar to (\ref{TotalLikelihood_eq}) and fit the
parameters of $\pi (.)$ by maximizing the likelihood.

~

\begin{figure}[t]
\centerline{\includegraphics{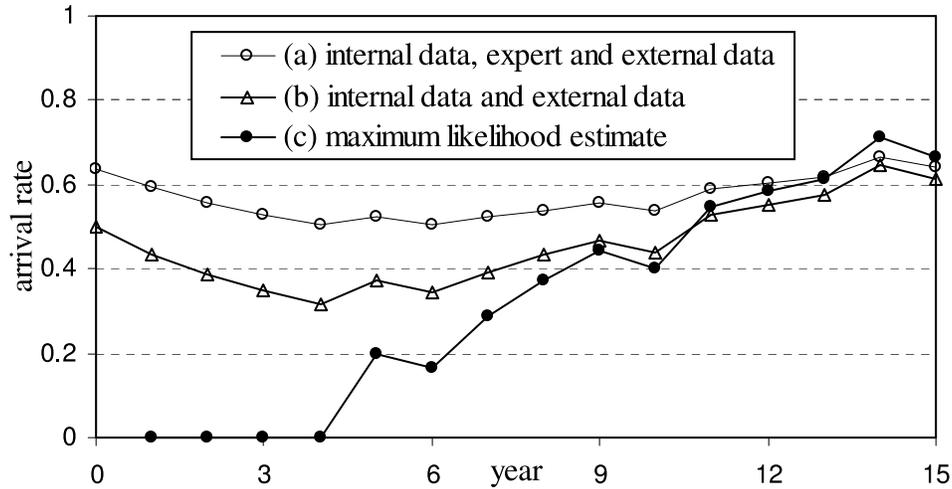}} \caption{Three estimators
for the Poisson arrival rate vs the observation year: (a) Bayesian
estimator combining internal data, expert opinion $\hat {\vartheta }
= 0.7$ and external data; (b) Bayesian estimator combining internal
data and external data; (c) MLE based on internal data only. The
internal data annual counts (0,0,0,0,1,0,1,1,1,0,2,1,1,2,0) were
sampled from the $Poisson(0.6)$. The prior implied by external data
is $Gamma(\alpha,\beta)$ with $\rm{mean}=0.5$.}
\label{CombiningData_fig}
\end{figure}

\noindent\textbf{Example.} Suppose that the annual frequency $N$ is
modeled by a Poisson distribution $Poisson(\lambda )$ and the prior
$\pi (\lambda )$ for $\lambda $ is $Gamma(\alpha ,\beta )$. As
described above, the prior can be estimated using either expert
opinions or external data. The expert may specify the ``best''
estimate for the expected number of events $E[E[N\vert \lambda ]] =
E[\lambda ]$ and an uncertainty that the ``true'' $\lambda $ for
next year is within the interval [$a$,$b$] with the probability $\Pr
[a \le \lambda \le b] = p$. Then the equations $E[\lambda ] = \alpha
\beta$ and $p = \int_a^b {\pi (\lambda )d\lambda } $ can be solved
numerically to estimate the structural parameters $\alpha $ and
$\beta $. In the insurance industry, the uncertainty for the
``true'' $\lambda $ is often measured in terms of the coefficient of
variation, $\rm{Vco}(\lambda ) = \sqrt {\rm{var}(\lambda )} /
E[\lambda ]$. Given the estimates for $E[\lambda ] = \alpha \beta $
and $\rm{Vco}(\lambda ) = 1 / \sqrt \alpha ,$ the structural
parameters $\alpha $ and $\beta $ are easily estimated. For example,
if the expert specifies (or external data imply) that $E[\lambda ] =
0.5$ and $\Pr [0.25 \le \lambda \le 0.75] = 2 / 3,$ then we can fit
a prior $Gamma(\alpha \approx 3.407, \beta \approx 0.147)$. This
prior is used in Figure \ref{CombiningData_fig}, presenting the
posterior best estimate for the arrival rate calculated using
(\ref{PosteriorMeanPoisson_eq}) and referred to as estimator (b),
when the annual counts data $N(k)$, $k = 1,\ldots,15$ are simulated
from $Poisson(0.6)$. Note that, in Figure \ref{CombiningData_fig},
the prior is considered to be implied by external data. On the same
Figure we show the standard MLE, $\hat {\lambda }_{^k}^{MLE} =
\frac{1}{k}\sum\nolimits_{i = 1}^k {n(i)} $, referred to as
estimator (c). For a small number of observed years the Bayesian
estimator (b) is more accurate as it takes prior information into
account. For a large sample size, both the MLE and Bayesian
estimators converge to the true value 0.6. Also, the Bayesian
estimator is more stable (smooth) with respect to bad years. The
same behavior is observed if the experiment is repeated many times
with different sequences of random numbers. This and other examples
can be found in \cite{ShWu06}.

\subsection{Combining three data sources}
\label{CombiningThreeSources_subsec} In Section
\ref{CombiningTwoSources_subsec}, Bayesian inference was used to
combine two data sources, i.e. expert opinion with internal data, or
external data with internal data. An approach to combine all three
data sources (internal data, expert opinion and external data) can
be accomplished as described in Lambrigger \textit{et al}.
\cite{LaShWu07}. Consider data ${\rm {\bf X}}$ and expert opinions
${{\bm \upsilon }}$ on parameter ${{\bm \theta }}$. Then the
posterior is

\begin{equation}
\label{PosteriorThreeSources_eq} \pi ({{\bm \theta }}\vert {\rm {\bf
x}},{{\bm \upsilon }}) \propto \ell _1 ({\rm {\bf x}}\vert { {\bm
\theta }})\ell _2 ({{\bm \upsilon }}\vert {{\bm \theta }})\pi ({{\bm
\theta }}),
\end{equation}

\noindent where $\ell _1 ({\rm {\bf x}}\vert {{\bm \theta }})$ is
the likelihood of data given ${{\bm \theta }}$, $\ell_2 ({{\bm
\upsilon }}\vert {{\bm \theta }})$ is the likelihood of expert
opinions and $\pi ({{\bm \theta }})$ is the prior density estimated
using external data. This posterior for ${{\bm \theta }}$ combines
information from internal data, expert opinions and external data.
Here it is assumed that given ${{\bm \theta }}$, expert opinions are
independent from internal data. A more general relation $\pi ({{\bm
\theta }}\vert {\rm {\bf x}},{{\bm \upsilon }}) \propto \ell ({\rm
{\bf x}},{{\bm \upsilon }}\vert {{\bm \theta }})\pi ({{\bm \theta
}})$ can be considered to avoid this assumption.

For illustration purposes, consider modeling of the annual counts:
assume that the annual counts $N(1),\ldots,N(T)$ are iid from
$Poisson(\lambda )$; expert opinions $\upsilon_1,\ldots,\upsilon_M$
on $\lambda $ are iid from $Gamma(\xi ,\lambda / \xi )$; and the
prior on $\lambda $ is $Gamma(\alpha ,\beta )$. Then the posterior
is the generalized inverse gamma density
\begin{eqnarray}
\label{PosteriorThreeSources_PoissonGamma_eq}
 \pi (\lambda \vert {\rm {\bf n}},{{\bm \upsilon }}) & \propto & \pi
(\lambda )\ell ({\rm {\bf n}}\vert \lambda )\ell ({{\bm \upsilon
}}\vert
\lambda ) \nonumber \\
 \quad &\propto& \lambda^{\alpha - 1}e^{ -
\lambda / \beta }\prod\limits_{t = 1}^T {e^{ - \lambda
}\frac{\lambda ^{n(t)}}{n(t)!}} \prod\limits_{m = 1}^M {e^{ -
\upsilon _m \xi / \lambda }\frac{\upsilon _m^{\xi - 1} }{(\lambda /
\xi )^\xi }} \propto \lambda
^{{\kern 1pt} \nu }e^{ - \lambda \omega - \phi / \lambda }, \nonumber \\
 \nu &=& \alpha - 1 + \sum\limits_{t = 1}^T {n(t)} - M\xi ,\quad \omega = T +
1 / \beta ,\quad \phi = \xi \sum\limits_{m = 1}^M {\upsilon _m }.
\end{eqnarray}

In Figure \ref{CombiningData_fig}, we show the posterior best
estimate for the arrival rate $E[\lambda \vert {\rm {\bf N}},{{\bm
\upsilon }}]$ combining three data sources (referred to as estimator
(a)) and compare it with the estimator $E[\lambda \vert {\rm {\bf
N}}]$ combining internal and external data (referred to as estimator
(b), also see (21)). The counts $N(k)$, $k = 1,\ldots,15$ are
simulated from $Poisson(0.6)$; the assumed prior distribution
implied by external data is the same as considered in the example in
Section \ref{EstimatingPrior_subsec}, i.e. $Gamma(\alpha \approx
3.41,\;\beta \approx 0.15)$ such that $E[\lambda ] = 0.5$ and $\Pr
[0.25 \le \lambda \le 0.75] = 2 / 3$; and there is one expert
opinion $\hat {\upsilon } = 0.7$ from the distribution with
${\rm{Vco}}(\upsilon \vert \lambda ) = $0.5, i.e $\xi = 4$. The
standard maximum likelihood estimate of the arrival rate $\lambda
_k^{MLE} = \frac{1}{k}\sum\nolimits_{i = 1}^k {n(i)} $ is referred
to as estimator (c). Estimator (a), combining all three data
sources, certainly outperforms other estimators and is more stable
around the true value, especially for small data sample size. All
estimators converge to the true value as the number of observed
years increases. The same behavior is observed if the experiment is
repeated; see detailed discussions in \cite{LaShWu07}.

\section{Modeling dependence}
\label{Dependence_sec} Basel II requires (see
\cite[p.152]{BaselII06}) that: ``\textit{Risk measures for different
operational risk estimates must be added for purposes of calculating
the regulatory minimum capital requirement. However, the bank may be
permitted to use internally determined correlations in operational
risk losses across individual operational risk estimates, provided
it can demonstrate to the satisfaction of the national supervisor
that its systems for determining correlations are sound, implemented
with integrity, and take into account the uncertainty surrounding
any such correlation estimates (particularly in periods of stress).
The bank must validate its correlation assumptions using appropriate
quantitative and qualitative techniques}''. Thus if dependence is
properly quantified between all risk cells $j = 1,\ldots,J$ then,
under the LDA model (\ref{LDAmodel_eq}), the capital is calculated
as

\begin{equation}
\label{totVaR_eq} VaR_{0.999} \left( {Z_{(\bullet )} (T + 1) =
\sum\limits_{j = 1}^J {Z_j (T + 1)} } \right),
\end{equation}

\noindent otherwise the capital should be estimated as

\begin{equation}
\label{VaRperfectDependence_eq} \sum\limits_{j = 1}^J {VaR_{0.999}
\left( {Z_j (T + 1)} \right)}.
\end{equation}

\noindent Adding up VaRs for capital estimation is equivalent to an
assumption of perfect positive dependence between the annual losses
$Z_j (T + 1)$, $j = 1,\ldots,J$. In principle, VaR can be estimated
at any level of granularity and then the capital is calculated as a
sum of resulting VaRs. Often banks quantify VaR for business lines
and add up these estimates to get capital, but for simplicity of
notations, (\ref{VaRperfectDependence_eq}) is given at the level of
risk cells. It is expected that the capital under (\ref{totVaR_eq})
is less than (\ref{VaRperfectDependence_eq}); 20{\%} diversification
is not uncommon. However, it is important to note that VaR is not a
coherent risk measure, see Artzner \textit{et al}.
\cite{ArDeEbHe99}. In particular, under some circumstances VaR
measure may fail a sub-additivity property

\begin{equation}
\label{Subadditivity_eq} VaR_q (Z_{(\bullet )} (T + 1)) \le
\sum\limits_{j = 1}^J {VaR_q (Z_j (T + 1)} ),
\end{equation}

\noindent see Embrechts \textit{et al}. \cite{EmNeWu07} and
\cite{EmLaWu09}, i.e. dependence modeling could also increase VaR.
As can be seen from the literature, the dependence between different
ORs can be introduced by:
\begin{itemize}

\item Modeling dependence between the annual counts via a copula, as
described in Frachot \textit{et al}. \cite{FrRoSa04}, Bee
\cite{Bee05a}, Aue and Klakbrener \cite{AuKl06};

\item Using common shock models to introduce events common across
different risks and leading to the dependence between frequencies
studied in Lindskog and McNeil \cite{LiMc03} and Powojowski
\textit{et al}. \cite{PoReTu02}. Dependence between severities
occurring at the same time is considered in Lindskog and McNeil
\cite{LiMc03};

\item Modeling dependence between the $k$th severities or between
$k$th event times of different risks; see Chavez-Demoulin \textit{et
al}. \cite{ChEmNe06} (e.g. $1^{st}$, $2^{nd}$, etc losses/event
times of the $j$th risk are correlated to the $1^{st}$, $2^{nd}$,
etc losses/event times of the $i$th risk respectively);

\item Modeling dependence between the annual losses of different
risks via copulas; see Giacometti \textit{et al}. \cite{GiRaChBe08},
Embrechts and Puccetti \cite{EmPu08};

\item Using the multivariate compound Poisson model based on L\'{e}vy
copulas suggested in B\"{o}cker and Kl\"{u}ppelberg \cite{BoKl08},
\cite{BoKl09};

\item Using structural models with common (systematic) factors that
can lead to the dependence between severities and frequencies of
different risks and within risk;

\item Modeling dependence between severities and frequencies from
different risks and within risk using dependence between risk
profiles considered in Peters \textit{et al}. \cite{PeShWu09}.

\end{itemize}

\noindent Below, we describe the main concepts involved into these
approaches.

\subsection{Copula}
\label{Copula_subsec} The concept of a copula is a flexible and
general technique to model dependence; for an introduction see e.g.
Joe \cite{Joe97} and Nelson \cite{Nelson99}; and for application in
financial risk management see e.g. \cite[Section 5]{McFrEm05}. In
brief, a copula is a $d$-dimensional multivariate distribution on
$[0,1]^d$ with uniform margins. Given a copula function $C$(.) and
univariate marginal distributions $F_1 (.),\ldots,F_d (.)$, the
joint distribution with these margins can be constructed as

\begin{equation}
\label{SklarTheorem_eq} F(x_1 ,\ldots,x_d )) = C(F_1 (x_1
),\ldots,F_d (x_d )).
\end{equation}

\noindent A well known theorem due to Sklar, published in 1959, says
that one can always find a unique copula $C$(.) for a joint
distribution with given continuous margins. In the case of discrete
distributions this copula may not be unique. The most commonly used
copula (due its simple calibration and simulation) is the Gaussian
copula, implied by the multivariate Normal distribution. It is a
distribution of $U_1 = F_N (X_1 ),\ldots,U_d = F_N (X_d )$, where
$F_N (.)$ is the standard Normal distribution and $X_1 ,\ldots,X_d $
are from the multivariate Normal distribution $F_\Sigma ($.) with
zero mean, unit variances and correlation matrix $\Sigma $.
Formally, in explicit form, the Gaussian copula is

\begin{equation}
\label{GaussianCopula_eq} C_\Sigma ^{Ga} (u_1 ,\ldots,u_n ) =
F_\Sigma (F_N^{ - 1} (u_1 ),\ldots,F_N^{ - 1} (u_d )).
\end{equation}

\noindent There are many other copulas (e.g. t-copula, Clayton
copula, Gumbel copulas to mention a few) studied in academic
research and used in practice, that can be found in the referenced
literature.

\subsection{Dependence between frequencies via copula}
\label{FreqDependence_subsec} 
\begin{figure}[t]
\centerline{\includegraphics{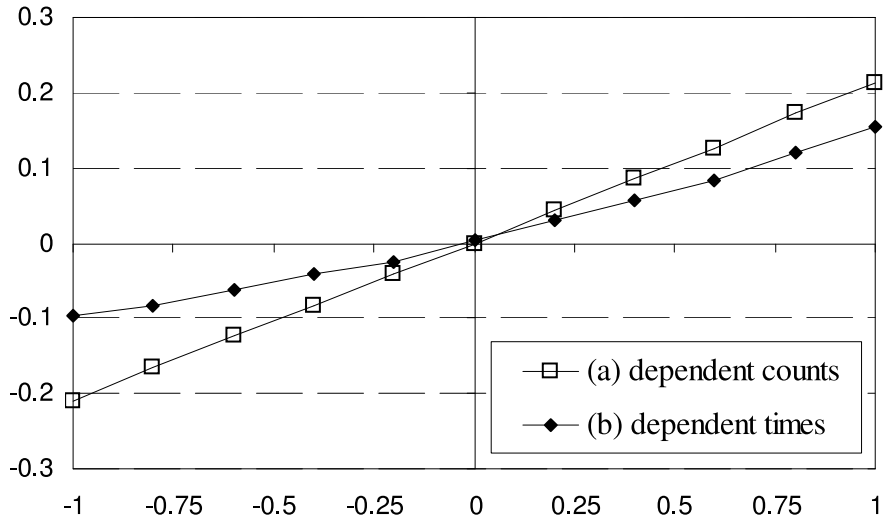}} \caption{Spearman's rank
correlation between the annual losses $\rho_S(Z_1,Z_2 )$ vs the
Gaussian copula parameter $\rho $: (a) -- copula between counts $N_1
$ and $N_2$; (b) -- copula between inter-arrival times of two
Poisson processes. Marginally, the frequencies are from $Poisson(5)$
and $Poisson(10)$ respectively and the severities are iid from
$LN(1,2)$ for both risks.} \label{FreqDependence_fig}
\end{figure}

The most popular approach in practice
is to consider a dependence between the annual counts of different
risks via a copula. Assuming a $J$-dimensional copula $C(.)$ and the
marginal distributions $P_j (.)$ for the annual counts $N_1
(t),\ldots,N_J (t)$ leads to a model

\begin{equation}
\label{FreqDependence_eq} N_1 (t) = P_1^{ - 1} (U_1 (t)),\ldots,N_J
(t) = P_J^{ - 1} (U_J (t)),
\end{equation}

\noindent where $U_1 (t),\ldots,U_J (t)$ are uniform (0,1) rvs from
a copula $C(.)$ and $P_j^{ - 1} (.)$ is the inverse marginal
distribution of the counts in the $j$th risk. Here, $t$ is a
discrete time (typically in annual units but shorter steps might be
needed to calibrate the model) and usually the counts are assumed to
be independent between different $t$ steps. The approach allows us
to model both positive and negative dependence between counts. As
reported in the literature, the implied dependence between annual
losses even for a perfect dependence between counts is relatively
small and as a result the impact on capital is small too. Some
theoretical reasons for the observation that frequency dependence
has only little impact on the operational risk capital charge are
given in \cite{BoKl08}. As an example, in Figure
\ref{FreqDependence_fig} we plot Spearman's rank correlation between
the annual losses of two risks, $Z_1 $ and $Z_2 $, induced by the
Gaussian copula dependence between frequencies. Marginally, the
frequencies $N_1 $ and $N_2 $ are from the Poisson distributions
with the intensities $\lambda = 5$ and $\lambda = 10$ respectively
and the severities are from $LN(\mu = 1,\sigma = 2)$ distributions
for both risks.

\subsection{Dependence between aggregated losses via copula}
\label{AggrLossDependence_subsec} Dependence between the aggregated
losses can be introduced similarly to (\ref{FreqDependence_eq}). In
this approach, one can model the aggregated losses as

\begin{equation}
\label{AggrLossDependence_eq} Z_1 (t) = F_1^{ - 1} (U_1
(t)),\ldots,Z_J (t) = F_J^{ - 1} (U_J (t)),
\end{equation}

\noindent where $U_1 (t),\ldots,U_J (t)$ are uniform (0,1) rvs from
a copula $C(.)$ and $F_j^{ - 1} (.)$ is the inverse marginal
distribution of the aggregated loss of the $j$-th risk. Note that
the marginal distribution $F_j (.)$ should be calculated using
frequency and severity distributions. Typically, the data are
available over several years only and a short time step $t$ (e.g.
quarterly) is needed to calibrate the model. This dependence
modeling approach is probably the most flexible in terms of the
range of achievable dependencies between risks; e.g. perfect
positive dependence between the annual losses is achievable.
However, note that this approach may create difficulties with
incorporation of insurance into the overall model. This is because
an insurance policy may apply to several risks with the cover limit
applied to the aggregated loss recovery; see Section
\ref{Insurance_sec}.

\subsection{Dependence between the kth event times/losses}
\label{KthLossTimesDependence_subsec} Theoretically, one can
introduce dependence between the $k$th severities or between the
$k$th event inter-arrival times or between the $k$th event times of
different risks. For example: $1^{st}$, $2^{nd}$, etc losses of the$
j$th risk are correlated to the $1^{st}$, $2^{nd}$, etc losses of
the $i$th risk respectively while the severities within each risk
are independent. The actual dependence can be done via a copula
similar to (\ref{FreqDependence_eq}), for an accurate description we
refer to \cite{ChEmNe06}. Here, we would like to note that a
physical interpretation of such models can be difficult. Also, an
example of dependence between annual losses induced by dependence
between the $k$th inter-arrival times is presented in Figure
\ref{FreqDependence_fig}.

\subsection{Modeling dependence via L\'{e}vy copulas}
\label{LevyCopula_subsec} B\"{o}cker and Kl\"{u}ppelberg
\cite{BoKl08,BoKl09} suggested to model dependence in frequency and
severity between different risks at the same time using a new
concept of L\'{e}vy copulas, see e.g. \cite[Sections
5.4-5.7]{CoTa04}. It is assumed that each risk follows to a
univariate compound Poisson process (that belongs to a class of
L\'{e}vy processes). Then, the idea is to introduce the dependence
between risks in such a way that any conjunction of different risks
constitutes a univariate compound Poisson process. It is achieved
using the multivariate compound Poisson processes based on L\'{e}vy
copulas. Note that, if dependence between frequencies or annual
losses is introduced via copula as in (\ref{FreqDependence_eq}) or
(\ref{AggrLossDependence_eq}), then the conjunction of risks does
not follow to a univariate compound Poisson.

The precise definitions of L\'{e}vy measure and L\'{e}vy copula are
beyond the purpose of this paper and can be found in the above
mentioned literature. Here, we would like to mention that in the
case of a compound Poisson process L\'{e}vy measure is the expected
number of losses per unit of time with a loss amount in a
pre-specified interval, $\bar {\Pi }_j (x) = \lambda _j \Pr (X_j >
x)$. Then the multivariate L\'{e}vy measure can be constructed from
the marginal measures and a L\'{e}vy copula $\tilde {C}$ as

\begin{equation} \bar {\Pi }(x_1 ,\ldots,x_d ) = \tilde {C}(\bar {\Pi }_1 (x_1
),\ldots,\bar {\Pi }_d (x_d ))
\end{equation}

\noindent which is somewhat similar to (\ref{SklarTheorem_eq}) in a
sense that the dependence structure between different risks can be
separated from the marginal processes. However, it is quite a
different concept. In particular, a L\'{e}vy copula for processes
with positive jumps is $[0,\infty )^d \to [0,\infty )$ mapping while
a standard copula (\ref{SklarTheorem_eq}) is $[0,1]^d \to [0,1]$
mapping. Also, a L\'{e}vy copula controls dependence between
frequencies and dependence between severities (from different risks)
at the same time. The interpretation of this model is that
dependence between different risks is due to the loss events
occurring at the same time. Important implication of this approach
is that a total bank's loss can be modeled as a compound Poisson
process with some intensity and iid severities. If this common
severity distribution is sub-exponential then closed-form
approximation (\ref{VaRtailApprox_eq}) can be used to estimate VaR
of the total annual loss in a bank.

\subsection{Structural model with common factors}
\label{CommonFactorsDependence_subsec} The use of common
(systematic) factors is useful to identify dependent risks and to
reduce the number of required correlation coefficients that must be
estimated, see e.g. \cite[Section 3.4]{McFrEm05}. Structural models
with common factors to model dependence are widely used in credit
risk, see industry examples in \cite[Section 8.3.3]{McFrEm05}. For
OR, these models are qualitatively discussed in Marshall
\cite[Sections 5.3 and 7.4]{Marshall01} and there are unpublished
examples of a practical implementation. As an example, assume a
Gaussian copula for the annual counts of different risks and
consider one common (systematic) factor $\Omega (t)$ affecting the
counts as follows:

\begin{eqnarray}
\label{StructDependence_eq}
 Y_j (t) &= &\rho _j \Omega (t) + \sqrt {1 - \rho _j^2 } W_j (t),\quad j =
1,\ldots,J;\quad \nonumber \\
 N_1 (t) &=& P_1^{ - 1} \left( {F_N (Y_1 (t))} \right),\ldots,N_J (t) = P_J^{ -
1} \left( {F_N (Y_J (t))} \right).
\end{eqnarray}

\noindent Here, $W_1 (t),\ldots,W_J (t)$ and $\Omega (t)$ are
independent rvs from the standard Normal distribution. All rvs are
independent between different time steps $t$. Given $\Omega (t)$,
the counts are independent but unconditionally the risk profiles are
dependent if the corresponding $\rho_j$ are nonzero. In this
example, one should identify $J$ correlation parameters $\rho_j$
only instead of $J(J-1)/2$ parameters of the full correlation
matrix.

Extension of this approach to many factors $\Omega _k (t)$, $k =
1,\ldots,K$ is easy:

\begin{equation}
\label{ManyFactorDependence_eq} Y_j (t) = \sum\limits_{k = 1}^K
{\rho _{jk} \Omega _k (t)} + \sqrt {1 - \sum\limits_{k = 1}^K {\rho
_{jk} \rho _{jm} {\rm{cov}}(\Omega _k (t)\Omega _m (t))} } W_j (t),
\end{equation}

\noindent where $\Omega _1 (t),\ldots,\Omega _K (t)$ are from a
multivariate Normal distribution with zero means, unit variances and
some correlation matrix. This approach can also be extended to
introduce a dependence between both severities and frequencies. For
example, in the case of one factor, one can structure the model as
follows:
\begin{eqnarray}
\label{FactorDpendenceFreqSev_eq}
 Y_j (t) &=& \rho _j \Omega (t) + \sqrt {1 - \rho _j^2 } W_j (t),\quad j =
1,\ldots,J; \nonumber\\
 N_j (t) &=& P_j^{- 1} \left( {F_N (Y_j (t))} \right),\quad j =
1,\ldots,J; \nonumber \\
 R_j^{(s)} (t) &=& \tilde {\rho }_j \Omega (t) + \sqrt {1 - \tilde {\rho }_j^2
} V_j^{(s)} (t), \quad s = 1,\ldots,N_j (t),\quad j = 1,\ldots,J; \nonumber \\
 X_j^{(s)} (t) &=& F_j^{ - 1} \left( {F_N (R_j^{(s)} (t))}
\right), \quad s = 1,\ldots,N_j (t),\quad j = 1,\ldots,J.\nonumber
\end{eqnarray}

\noindent Here: $W_j (t)$, $V_j^{(s)} (t)$, $s = 1,\ldots,N_j (t)$,
$j = 1,\ldots,J$ and $\Omega (t)$ are iid from the standard Normal
distribution. Again, the logic is that there is a factor affecting
severities and frequencies within a year such that conditional on
this factor, severities and frequencies are independent. The factor
is changing stochastically from year to year, so that
unconditionally there is dependence between frequencies and
severities. Also note that in such setup, there is a dependence
between severities within a risk category. Often, common factors are
unobservable and practitioners use generic intuitive definitions
such as: changes in political, legal and regulatory environments,
economy, technology, system security, system automation, etc.
Several external and internal factors are typically considered, so
that some of the factors affect frequencies only (e.g. system
automation), some factors affect severities only (e.g. changes in
legal environment) and some factors affect both the frequencies and
severities (e.g. system security).

\subsection{Stochastic and dependent risk profiles}
\label{ProfileDependence_subsec} Consider the LDA for risk cells $j
= 1,\ldots,J$:

\begin{equation}
Z_j (t) = \sum\limits_{s = 1}^{N_j (t)} {X_j^{(s)} (t)} ,\quad t =
\mbox{1,2,\ldots} \quad ,
\end{equation}

\noindent where $N_j (t)\sim P_j (.\vert \lambda _t^{(j)} )$ and
$X_j^{(s)} (t)\sim F_j (.\vert \psi _t^{(j)} )$. Hereafter, notation
$X\sim F(.)$ means that $X$ is a rv from distribution $F(.)$. It is
realistic to consider that the risk profiles ${{\bm \lambda}}_t =
(\lambda_t^{(1)} ,\ldots,\lambda_t^{(J)} )$ and ${{\bm \psi }}_t =
(\psi _t^{(1)} ,\ldots,\psi _t^{(J)})$ are not constant but changing
in time stochastically due to changing risk factors (e.g. changes
business environment, politics, regulations, etc). Also it is
realistic to say that risk factors affect many risk cells and thus
the risk profiles are dependent. One can model this by assuming some
copula $C(.)$ and marginal distributions for the risk profiles (also
see \cite{PeShWu09a}), i.e. the joint distribution between the risk
profiles is

\begin{equation}
\label{RiskProfileCopula_eq} F\left( {{{\bm\lambda}}(t),{{\bm \psi
}}(t)} \right) = C\left( {G_1 (\lambda _1 (t)),\ldots,G_J (\lambda
_J (t)),H_1 (\psi _1 (t)),\ldots,H_J (\psi_J (t))} \right),
\end{equation}

\noindent where $G_j (.)$ and $H_j (.)$ are the marginal
distributions of $\lambda _j (t)$ and $\psi _j (t)$ respectively.
Dependence between the risk profiles will induce a dependence
between the annual losses. This general model can be used to model
dependence between the annual counts; between the severities of
different risks; between the severities within a risk; and between
the frequencies and severities. The likelihood of data (counts and
severities) can be derived but involves a multidimensional integral
with respect to latent variables (risk profiles). Advanced MCMC
methods (such as the Slice Sampler method used in \cite{PeShWu09a})
can be used to fit the model. For example, consider the bivariate
case ($J = 2)$ where:
\begin{itemize}
\item Frequencies $N_j (t)\sim Poisson(\lambda _j (t))$ and severities
$X_j^{(s)}(t)\sim LN(\mu _j (t),\sigma _j (t))$;

\item
$\lambda _1 (t)\sim Gamma(2.5,2)$, $\lambda _2 (t)\sim Gamma(5,2)$,
$\mu _j (t)\sim Normal(1,1)$, $\sigma _j (t) = 2;$

\item The dependence between $\lambda _1 (t)$, $\lambda _2 (t)$, $\mu
_1 (t)$ and $\mu _2 (t)$ is a Gaussian copula.

\end{itemize}

\begin{figure}[b]
\centerline{\includegraphics{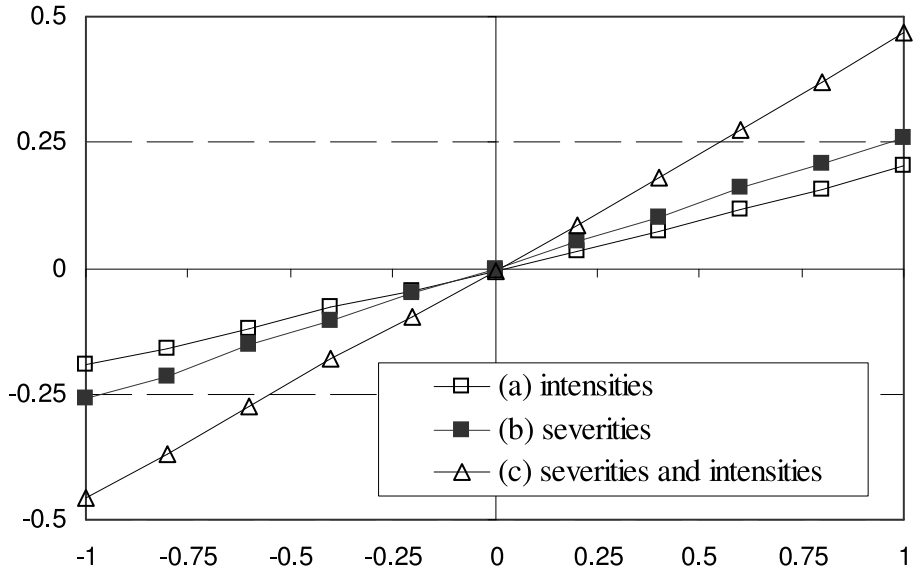}} \caption{Spearman's rank
correlation $\rho_{S}(Z_1,Z_2)$ between annual losses vs the
Gaussian copula parameter $\rho$: (a) -- copula for the frequency
profiles $\lambda _1 $ and $\lambda _2 $; (b) -- copula for the
severity profiles $\mu_1 $ and $\mu _2 $; (c) -- copula for
$\lambda_1 $ and $\lambda_2 $ and the same copula for $\mu_1 $ and
$\mu_2$} \label{DependenceBetweenProfiles_fig}
\end{figure}

\noindent Figure \ref{DependenceBetweenProfiles_fig} shows the
induced dependence between the annual losses $Z_1 (t)$ and $Z_2 (t)$
vs the copula dependence parameter for three cases: if only $\lambda
_1 (t)$ and $\lambda _2 (t)$ are dependent; if only $\mu _1 (t)$ and
$\mu _2 (t)$ are dependent; if the dependence between $\lambda _1
(t)$ and $\lambda _2 (t)$ is the same as between $\mu _1 (t)$ and
$\mu _2 (t)$. In all cases the dependence is Gaussian copula.

\subsection{Common shock processes}
\label{CommonShockDependence_subsec} Modeling OR events affecting
many risk cells can be done using common shock process models; see
Johnson \textit{et al}. \cite[Section 37]{JoKoBa97}. In particular,
consider $J$ risks with the event counts $N_j (t) = N^{(C)}(t) +
\tilde {N}_j (t)$, where $\tilde {N}_j (t)$, $j = 1,\ldots,J$ and
$N^{(C)}(t)$ are generated by independent Poisson processes with
intensities $\tilde {\lambda }_j $ and $\lambda _C $ respectively.
Then $N_j (t)$, $j = 1,\ldots,J$ are Poisson distributed with
intensities $\lambda _j = \tilde {\lambda }_j + \lambda_C$
marginally and are dependent via the common events $N^{(C)}(t)$. The
linear correlation and covariance between risk counts are $\rho (N_i
(t),N_j (t)) = \lambda _C / \sqrt {\lambda _i \lambda _j } $ and
${\rm{cov}}(N_i (t),N_j (t)) = \lambda _C $ respectively. Only a
positive dependence between counts can be modeled using this
approach. Also, note that the covariance for any pair of risks is
the same though the correlations are different. More flexible
dependence can be achieved by allowing a common shock process to
contribute to the $k$-th risk process with some probability $p_k $;
then ${\rm{cov}}(N_i (t),N_j (t)) = \lambda _C p_i p_j $. This
method can be generalized to many common shock processes; see
\cite{LiMc03} and \cite{PoReTu02}. It is also reasonable to consider
the dependence between the severities in different risk cells that
occurred due to the same common shock event.

\section{Insurance}
\label{Insurance_sec}
Many ORs are insured. If a loss occurred and
it is covered by an insurance policy then part of the loss will be
recovered. A typical policy will provide a recovery $R$ for a loss
$X$ subject to the excess amount (deductible) $D$ and top cover
limit amount $U$ as follows:

\begin{equation}
\label{InsuranceRecovery_eq} R = \left\{ {{\begin{array}{*{20}c}
 {0,\quad \mbox{if }0 \le X < D;} \hfill \\
 {X - D,\quad \mbox{if }D \le X < U + D;} \hfill \\
 {U,\quad \mbox{if }D + U \le X.} \hfill \\
\end{array} }} \right.
\end{equation}

\noindent That is the recovery will take place if the loss is larger
than the excess and the maximum recovery that can be obtained from
the policy is $U$. Note that in (\ref{InsuranceRecovery_eq}), the
time of the event is not involved and the top cover limit applies
for a recovery per risk event, i.e. for each event the obtained
recovery is subject of the top cover limit. Including insurance into
the LDA is simple; the loss severity in (\ref{LDAmodel_eq}) should
be simply reduced by the amount of recovery
(\ref{InsuranceRecovery_eq}) and can be viewed as a simple
transformation of the severity. However, there are several
difficulties in practice. Policies may cover several different risks
and different policies may cover the same risk. The top cover limit
may apply for the aggregated recovery over many events of one or
several risks (e.g. the policy will pay the recovery for losses
until the top cover limit is reached by accumulated recovery). These
aspects and special insurance haircuts required by Basel II
\cite{BaselII06} make recovery dependent on time. Accurate modeling
insurance accounting for practical details requires modeling the
event times rather than the annual counts only, e.g. a Poisson
process can be used to model the event times. It is not difficult to
incorporate the insurance into an overall model if a Monte Carlo
method is used to quantify the annual loss distributions.

The Basel II requires that the total capital reduction due to the
insurance recoveries is capped by 20{\%}. Incorporating insurance
into the LDA is not only important for capital reduction but also
beneficial for negotiating a fair premium with the insurer because
the distribution of the recoveries and its characteristics can be
estimated.

\section{Capital charge via full predictive distribution}
\label{CapitalViaPredDF_sec} Consider the annual loss in a bank (or
the annual loss at a different level depending on where the 0.999
quantiles are quantified; see Section \ref{Dependence_sec}) over the
next year, $Z(T + 1)$. Denote the density of the annual loss,
conditional on parameters ${{\bm \theta }}$, as $f(z(T + 1)\vert
{{\bm \theta }})$. Typically, given observations, the MLEs
${{\hat{\bm\theta }}}$ are used as the ``best fit'' point estimators
for ${{\bm \theta }}$. Then the annual loss distribution for the
next year is estimated as $f(z(T + 1)\vert {{\hat {\bm\theta }}})$
and its 0.999 quantile, $Q_{0.999}({{\hat {\bm\theta }}})$, is used
for the capital charge calculation.

However, the parameters ${{\bm \theta }}$ are unknown and it is
important to account for this uncertainty when capital charge is
estimated (especially for risks with small datasets) as discussed in
\cite{Shevchenko08a}. If Bayesian inference is used to quantify the
parameters through their posterior distribution $\pi ({{\bm \theta
}}\vert {\rm {\bf y}})$, then the full predictive density
(accounting for parameter uncertainty) of $Z(T + 1)$, given all data
${\rm {\bf Y}}$ used in the estimation procedure, is

\begin{equation}
\label{FullPredPDF_eq} f(z(T + 1)\vert {\rm {\bf y}}) = \int {f(z(T
+ 1)\vert {{\bm\theta }})\times \pi ({{\bm\theta }}\vert {\rm {\bf
y}})d{{\bm\theta }}} .
\end{equation}

\noindent Here, it is assume that, given parameters ${{\bm \theta
}}$, $Z(T + 1)$ and ${\rm {\bf Y}}$ are independent. If a
frequentist approach is taken to estimate the parameters, then ${
{\bm\theta }}$ should be replaced with ${{\hat{\bm\theta }}}$ and
the integration should be done with respect to the density of
parameter estimators ${{\hat{\bm\theta }}}$. The 0.999 quantile of
the full predictive distribution (\ref{FullPredPDF_eq}),

\begin{equation}
\label{QuantileFullPred_eq} Q_q^B = F_{Z(T + 1)\vert {\rm {\bf
Y}}}^{ - 1} (q) = \inf \{z:\Pr [Z(T + 1)
> z\vert {\rm {\bf Y}}] \le 1 - q\},\;q = 0.999,
\end{equation}

\noindent can be used as a risk measure for capital calculations.

Another approach under a Bayesian framework to account for parameter
uncertainty is to consider a quantile $Q_{0.999} ({{\bm \theta }})$
of the conditional annual loss density $f(.\vert {{\bm \theta }})$:

\begin{equation}
\label{DistrOfQuantile_eq} Q_q ({{\bm \theta }}) = F_{Z(T + 1)\vert
{{\bm \theta }}}^{ - 1} (q) = \inf \{z:\Pr [Z(T + 1) > z\vert {{\bm
\theta }}] \le 1 - q\},\;q = 0.999.
\end{equation}

\noindent Then, given that ${{\bm \theta }}$ is distributed as $\pi
({{\bm \theta }}\vert {\rm {\bf y}})$, one can find the distribution
of $Q_{0.999} ({{\bm \theta }})$ and form a predictive interval to
contain the true value with some probability. This is similar to
forming a confidence interval in the frequentist approach using the
distribution of $Q_{0.999} ({{\hat{\bm\theta }}})$, where
${{\hat{\bm\theta }}}$ is treated as random (usually, the Gaussian
approximation (\ref{InformationMatrix_eq}) is assumed for
${{\hat{\bm\theta }}})$. Often, if derivatives can be calculated
efficiently, the variance of $Q_{0.999} ({{\hat {\bm\theta }}})$ is
simply estimated via an error propagation method and a first order
Taylor expansion). Here, one can use deterministic algorithms such
as FFT or Panjer recursion to calculate $Q_{0.999} ({{\bm \theta
}})$ efficiently. Under this approach, one can argue that the
conservative estimate of the capital charge accounting for parameter
uncertainty should be based on the upper bound of the constructed
interval. Note that specification of the confidence level is
required and it might be difficult to argue that the commonly used
confidence level $0.95$ is good enough for estimation of the 0.999
quantile.

In OR, it seems that the objective should be to estimate the full
predictive distribution (\ref{FullPredPDF_eq}) for the annual loss
$Z(T + 1)$ over next year conditional on all available information
and then estimate the capital charge as a quantile $Q_{0.999}^B $ of
this distribution (\ref{QuantileFullPred_eq}.

Consider a risk cell in the bank. Assume that the frequency
$p(.\vert {{\bm\alpha }})$ and severity $f(.\vert {{\bm \beta }})$
densities for the cell are chosen. Also, suppose that the posterior
distribution $\pi ({{\bm\theta }}\vert {\rm {\bf y}})$, ${{\bm
\theta }} = ({{\bm \alpha }},{{\bm \beta }})$ is estimated. Then,
the full predictive annual loss distribution (\ref{FullPredPDF_eq})
in the cell can be calculated using Monte Carlo procedure with the
following logical steps:

%\noindent\hrulefill
\begin{enumerate}
\item For $k=1,\ldots,K$
\begin{enumerate}
\item For a given risk simulate the risk parameters ${{\bm \theta
}} = ({{\bm \alpha }},{{\bm \beta }})$ from the posterior
 $\pi ({{\bm \theta }}\vert {\rm {\bf y}})$. If the
posterior is not known in closed form then this simulation can be
done using MCMC (see Section \ref{BayesianApproach_subsec}). For
example, one can run MCMC for $K$ iterations beforehand and simply
take the $k$th iteration parameter values.

\item Given ${{\bm \theta }} = ({{\bm \alpha }},{{\bm
\beta }})$, simulate the annual number of events $N$ from $p(.\vert
{{\bm \alpha }})$ and severities $X^{(1)},\ldots,X^{(N)}$ from
$f(.\vert {{\bm \beta }})$; and calculate the annual loss $Z^{(k)} =
\sum\nolimits_{n = 1}^N {X^{(n)}} $.
\end{enumerate}
\item Next $k$
\end{enumerate}
%\noindent\hrulefill

\noindent Obtained annual losses $Z^{(1)},\ldots,Z^{(K)}$ are
samples from the full predictive density (\ref{FullPredPDF_eq}).
Extending the above procedure to the case of many risks is easy but
requires specification of the dependence model, see Section
\ref{Dependence_sec}. In this case, in general, all model parameters
(including the dependence parameters) should be simulated from their
joint posterior in Step 1. Then, given these parameters, Step 2
should simulate all risks with a chosen dependence structure. In
general, sampling from the joint posterior of all model parameters
can be accomplished via MCMC, see e.g. \cite{PeShWu09a,Dalla09}. The
0.999 quantile $Q_{0.999}^B$ and other distribution characteristics
can be estimated using the simulated samples in the usual way, see
Section \ref{compDistr_MC_sec}.

Note that in the above Monte Carlo procedure the risk profiles
${{\bm \alpha }}$ and ${{\bm \beta }}$ are simulated from their
posterior distribution for each simulation. Thus, we model both the
process risk (process uncertainty), which comes from the fact that
frequencies and severities are rvs, and the parameter risk
(parameter uncertainty), which comes from the fact that we do not
know the true values of ${{\bm \theta }} = ({{\bm \alpha }},{{\bm
\beta }})$. To calculate the conditional density $f(z\vert {\rm
{\hat{\bm\theta }}})$ and its quantile $Q_{0.999} ({{\hat {\bm\theta
}}})$ using parameter point estimators ${{\hat{\bm\theta }}}$, step
1 in the above procedure should be simply modified by setting ${{\bm
\theta }} = {{\hat{\bm\theta }}}$ for all simulations $k =
1,\ldots,K$. Thus, Monte Carlo calculations of $Q_{0.999}^B $ and
$Q_{0.999} ({{\hat {\bm\theta }}})$ are similar, given that
$\pi({{\bm \theta }}\vert {\rm {\bf y}})$ is known. If $\pi({{\bm
\theta }}\vert {\rm {\bf y}})$ is not known in closed form then it
can be estimated efficiently using Gaussian approximation or
available MCMC algorithms; see Section
\ref{BayesianApproach_subsec}.

\begin{figure}[t]
\centerline{\includegraphics{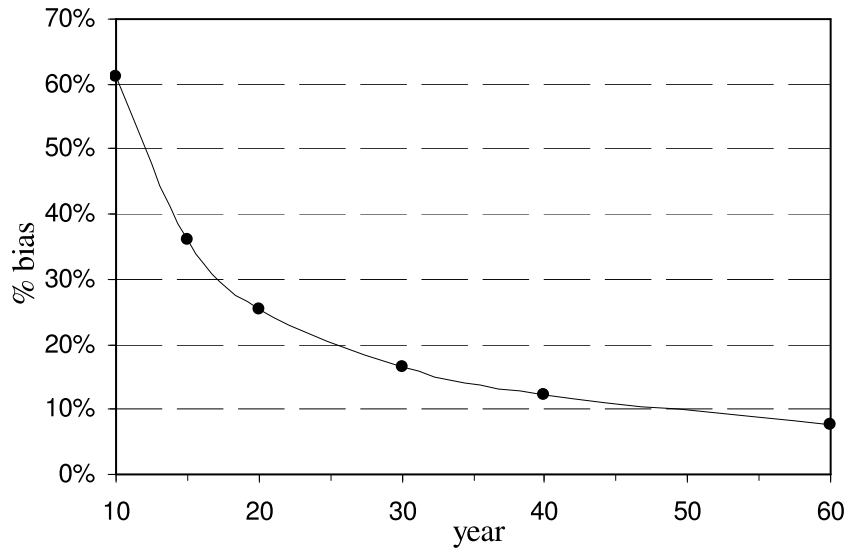}} \caption{Relative bias
(average over 100 realizations) in the 0.999 quantile of the annual
loss induced by the parameter uncertainty vs the number of
observation years. Losses were simulated from $Poisson(10)$ and
$LN(1,2)$.} \label{CapitalParamUncertainty_fig}
\end{figure}

The parameter uncertainty is ignored by the estimator $Q_{0.999}
({\hat {\bm\theta }})$ but is taken into account by $Q_{0.999}^B$.
Figure \ref{CapitalParamUncertainty_fig} presents results for the
relative bias (averaged over 100 realizations) $E[Q_{0.999}^B -
Q_{0.999} ({{\hat {\bm\theta }}})] / Q^{(0)}$, where ${{\hat
{\bm\theta }}}$ is MLE, $Q^{(0)}$ is the quantile of $f(.\vert {{\bm
\theta }}_0 )$ and ${{\bm \theta }}_0 $ is the true value of the
parameter. The frequencies and severities are simulated from
$Poisson(\lambda _0 = 10)$ and $LN(\mu _0 = 1,\sigma _0 = 2)$
respectively. Also, constant priors are used for the parameters so
that there are closed form expressions for the posterior. In this
example, the bias induced by parameter uncertainty is large: it is
approximately 10{\%} after 40 years (i.e. approximately 400 data
points) and converges to zero as the number of losses increases.

The parameter values used in the example may not be typical for some
ORs. One should do the above analysis with real data to find the
impact of parameter uncertainty. For example a similar analysis for
a multivariate case was performed in \cite{Dalla09} with real data.
For high frequency low impact risks, where a large amount of data is
available, the impact is certainly expected to be small. However for
low frequency high impact risks, where the data are very limited,
the impact can be significant. Also, see Mignola and Ugoccioni
\cite{MiUg06a} for discussion of uncertainties involved in OR
estimation.

\section{Conclusions}
In this paper we reviewed some methods suggested in the literature
for the LDA implementation. We emphasized that Bayesian methods can
be well suited for modeling OR. In particular, Bayesian framework is
convenient to combine different data sources (internal data,
external data and expert opinions) and to account for the relevant
uncertainties. Accurate quantification of the dependences between
ORs is a difficult task with many challenges to be resolved. There
are many aspects of the LDA that may require sophisticated
statistical methods and different approaches are hotly debated.

\section{Acknowledgments}
The author would like to thank Mario W\"{u}thrich, Hans
B\"{u}hlmann, Gareth Peters, Xiaolin Luo, John Donnelly and Mark
Westcott for fruitful discussions, useful comments and
encouragement. Also, the comments from anonymous referee helped to
improve the paper.

\renewcommand{\baselinestretch}{0.1}

%\bibliography{bibliography}
%\bibliographystyle{wileyj}

\renewcommand{\baselinestretch}{1.3}

\newpage

\begin{table}[htbp]
\begin{tabular}
{p{0.45\textwidth}p{0.505\textwidth}} \toprule \textbf{\quad Basel
II business lines (BL)}&
\textbf{\quad Basel II event types (ET)} \\
\cmidrule(r){1-1}\cmidrule(l){2-2}
\begin{itemize}
\item Corporate finance ($\beta_{1}=0.18$)
\item Trading {\&} Sales ($\beta_{2}=0.18$)
\item Retail banking ($\beta_{3}=0.12$)
\item Commercial banking ($\beta_{4}=0.15$)
\item Payment {\&} Settlement ($\beta_{5}=0.18$)
\item Agency Services ($\beta_{6}=0.15$)
\item Asset management ($\beta_{7}=0.12$)
\item Retail brokerage ($\beta_{8}=0.12$)
\end{itemize}
&
\begin{itemize}
\item Internal fraud
\item External fraud
\item{Employment practices and \par workplace safety}
\item Clients, products and business practices
\item Damage to physical assets
\item Business disruption and system failures
\item{Execution, delivery and \par process management}
\end{itemize}\\
\bottomrule
\end{tabular}
\caption{Basel II business lines and event types. $\beta _1
,\ldots,\beta _8$ are the business line factors used in the Basel II
Standardised Approach.} \label{BURT_table}
\end{table}

\bigskip

\pagebreak

\pagebreak

\end{document}